\documentclass[12pt,preprint]{aastex}
\begin{document}
\title{BROAD LINE EMISSION IN LOW-METALLICITY BLUE COMPACT DWARF GALAXIES:
EVIDENCE FOR STELLAR WIND, SUPERNOVA AND POSSIBLE AGN ACTIVITY}
\author{Yuri I. Izotov}
\affil{Main Astronomical Observatory, Ukrainian National Academy of Sciences,
27 Zabolotnoho str., Kyiv 03680, Ukraine}
\email{izotov@mao.kiev.ua}
\author{Trinh X. Thuan}
\affil{Astronomy Department, University of Virginia, P.O. Box 400325, 
Charlottesville, VA 22904-4325}
\email{txt@virginia.edu}
\and
\author{Natalia N. Guseva}
\affil{Main Astronomical Observatory, Ukrainian National Academy of Sciences,
27 Zabolotnoho str., Kyiv 03680, Ukraine}
\email{guseva@mao.kiev.ua}

\begin{abstract}
We present spectra of a large sample of low-metallicity blue compact dwarf
galaxies which exhibit broad components in their strong emission lines, mainly
in H$\beta$, [O {\sc iii}] $\lambda$4959, 5007 and H$\alpha$.
Twenty-three spectra have been obtained with the MMT\footnote{The MMT is operated by 
the MMT Observatory (MMTO), a joint venture of the Smithsonian Institution  
and the University of Arizona.}, 14 of which show broad 
emission. The remaining 21 spectra with broad emission have been 
selected from the Data Release 5 of the Sloan Digital Sky Survey.
The most plausible origin of broad
line emission is the evolution of massive stars and their
interaction with the circumstellar and interstellar medium. 
The broad emission with the lowest H$\alpha$ luminosities (10$^{36}$--10$^{39}$
erg s$^{-1}$) is likely produced in circumstellar envelopes
around hot Ofp/WN9 and/or LBV stars.
The broad emission with the highest H$\alpha$ luminosities 
(10$^{40}$--10$^{42}$ erg s$^{-1}$) probably arises from 
type IIp or type IIn supernovae (SNe). It can also come from
active galactic nuclei (AGN) containing intermediate-mass black holes, 
  although we find no strong evidence for hard non-thermal radiation in our 
sample galaxies. 
 The oxygen abundance in the host galaxies with SN candidates  
 is low and varies in the range 12 + log O/H = 7.36 -- 8.31.
However, type IIn SN / AGN candidates are found only in galaxies with 
12 + log O/H $\la$ 7.99.
Spectroscopic monitoring of these type IIn SN / AGN candidates 
over a time scale of several years is
necessary to distinguish between the two possibilities.
\end{abstract}


\section{INTRODUCTION}

One of the most remarkable features of blue compact dwarf (BCD) 
galaxies is their
high star formation rate, believed to occur in episodes lasting
only about 10 Myr, separated by longer quiescent 
periods of several Gyr \citep{fanelli88}.
During these starburst episodes, BCDs are characterized by
blue continua and strong narrow emission lines, evidence for a
large population of hot massive stars. The total number of O stars
responsible for the ionization of the interstellar medium in BCDs can
reach 10$^2$--10$^5$ \citep{G00}. Such a large number of massive stars can
ionize some 10$^5$ to 10$^7$ M$_\odot$ of interstellar gas. 
These massive stars can have a 
significant influence on the dynamics of the 
gas. Thus, stellar winds and supernovae (SNe) events 
can produce large gaseous velocities of 
$\ga$ 1000 km s$^{-1}$ which manifest themselves as broad 
emission lines. 
Detections of broad emission in the
H$\alpha$ line, characteristic of type IIp SNe, have been reported in 
BCDs by \citet{P97} and \citet{K98}.
More than 200 galaxies, most of which classified as BCDs, are known to exhibit 
broad emission features characteristic of Wolf-Rayet (WR) 
stars \citep{S99,G00,Z07}. 
They are called ``Wolf-Rayet galaxies''. In these WR galaxies,  
the broad permitted emission lines of some ions, most often
He {\sc ii}, N {\sc iii} and C {\sc iv}, are thought to originate
in the envelopes of massive stars
which are undergoing rapid mass loss.

While broad WR emission comes from stellar envelopes, the hydrogen
and forbidden lines in the spectra 
of some giant H {\sc ii} regions
and BCDs also exhibit broad low-intensity components 
\citep{D87,C90,R92,I96,I06}. These are likely to originate not only in the  
circumstellar envelopes of stars with stellar winds, but also 
in the H {\sc ii} regions 
themselves.  The full widths at zero intensity
(FWZI) of the low-intensity broad components can reach values of $\sim$
40--60\AA, which correspond to expansion velocities of about 2000 -- 3000
km s$^{-1}$, close to the terminal velocities of stellar winds from WR
and Of stars. In some BCDs, the regions of broad emission are associated with 
relatively strong 
narrow high-ionization emission lines of He {\sc ii}, [Ne {\sc v}],
[Fe {\sc v}] -- [Fe {\sc vii}] ions, with ionization potentials above 4 Ryd
\citep{Fr01,I01,I04a,I04b,TI05}. \citet{I04a} and \citet{TI05} 
have analyzed different mechanisms
for high-ionization line emission and have concluded that the most likely 
one appears to be fast shocks in very low-metallicity and 
dense interstellar media, 
triggered by single or multiple SNe events and/or stellar winds.

Theoretical studies of the interaction of large OB associations with
the interstellar medium \citep{W77,MK87,MM88} predict that multiple 
SN events can create a large and low-density stellar wind bubble which 
pushes out a supershell of denser ionized material. 
In disk galaxies,
under some conditions, the bubble 
can reach velocities of $\sim$10$^3$ km\ s$^{-1}$ or greater
 and result in a blowout. 
One of the objects where such a mechanism has been invoked 
to explain the broad emission is 
the star-forming region NGC 2363 $\equiv$ Mrk 71 in the dwarf irregular 
galaxy NGC 2366. It  
shows broad emission (FWHM $\sim$ 40 \AA), which corresponds to a 
velocity of 2400 km s$^{-1}$, in the  
H$\beta$, H$\alpha$, [O {\sc iii}] $\lambda$4959, 5007 lines. It  
has been studied in detail by \citet{R92} who found the 
broad emission to extend over a large region of space, $\ga$ 500 pc in size.
These authors have
considered several mechanisms which may explain  
the large-scale fast motion of the ionized gas. They 
concluded 
 that only a superbubble blowout is capable of accelerating
interstellar gas over such a large volume. However, all is not well 
with the superbubble model: such a blowout 
appears to be inconsistent with both
the observed lack of a hot cavity in NGC 2363 and the observed radius of the
expanding [O {\sc iii}] gaseous region which is smaller than the galaxy's 
scale height.

Another mechanism for broad emission in BCDs may be 
  active galactic nuclei (AGN) containing intermediate-mass black holes 
($M$ $\sim$ 10$^3$ -- 10$^5$ M$_\odot$). 
A large number of studies have been concerned with the 
properties of the broad line regions (BLR) and the narrow line
  regions (NLR) in AGN with solar and super-solar metallicities
\citep[e.g., ][]{FS85,FS89,Ho95,Ho97a,Ho97b,Ho03,K99}. When placed 
in emission-line diagnostic diagrams,
these AGN occupy a region well separated from that of 
star-forming galaxies 
\citep[e.g., ][]{K03}. 
The separation is not as clear, however, for low-metallicity AGN,
should they exist. 
Using photoionization models, \citet{Gr06} and \citet{S06} have shown that 
low-metallicity AGN occupy overlapping 
regions with those of normal H {\sc ii} regions 
in emission-line diagnostic diagrams, 
making it difficult to use these diagrams to distinguish non-thermal 
from thermal sources of
 ionization. 
To make the situation worse, 
the broad H$\alpha$ luminosities in AGN can be
 as low as 10$^{38}$ erg s$^{-1}$ \citep[e.g., ][]{Ho03}, which 
is just in the range of typical H$\alpha$ luminosities of SNe and 
stellar winds. Finally, spectra of SNe can mimic spectra of Sy
1 galaxies \citep{F89}. Therefore, additional evidence have to be
invoked to distinguish between low-metallicity AGN and star-forming
galaxies, e.g. the presence of high-ionization ions, the temporal evolution
of spectra, the spatial distribution of broad emission, etc.
To date, there is no confirmed case of a low-metallicity AGN.

In recent years, we have used the 6.5 meter MMT to obtain new high 
signal-to-noise ratio spectroscopic observations for a relatively 
large sample of BCDs, in the course of 
studying various problems such as the search  
of weak high-ionization emission lines in BCDs \citep{TI05} or 
the determination of the electron temperature from the Balmer 
jump \citep{G06}. These high signal-to-noise ratio spectra form an 
excellent sample to search for low-intensity broad line emission.
 We will hereafter refer to this sample as the ``MMT sample''. 
Additionally, a large spectroscopic data base has become available 
with the public release of the Sloan Digital Sky Survey
(SDSS). Its Data Release 5 (DR5) contains the spectra of 
some 650,000 galaxies \citep{A07}. We can also use  
this enormous data base to search for emission-line galaxies 
with broad emission. While the SDSS spectra have a lower signal-to-noise 
ratio than the MMT spectra, they make up for it by their vastly larger number. 
The second sample will be referred to as the ``SDSS sample''.
Together, the MMT and SDSS objects form a large enough sample 
of BCDs with broad emission to allow us to study 
the main physical mechanisms responsible for that emission.

We describe the MMT and SDSS samples in \S2. 
We discuss the main mechanisms for broad emission in \S3. 
In \S4, we consider whether low-metallicity AGN can exist. Our conclusions 
are summarized in \S5.

\section{THE DATA}

\subsection{The MMT sample}

The MMT sample is composed of all objects observed by \citet{TI05}. 
We have included in it not only objects with broad emission, but 
for comparison, also those BCDs that do not show it.  
 The galaxies in the MMT sample span a wide
range of oxygen abundances, 12+logO/H $\sim$ 7.1 -- 8.3, 
and their H {\sc ii}
regions are characterized by different ages and luminosities.
The detailed descriptions of the observations and 
of the data reduction are given in \citet{TI05}. 
Briefly, all observations were made with the Blue Channel of the MMT 
spectrograph on the nights of 2004 February 19 -- 20 and 2005 February 4. 
A 2\arcsec$\times$300\arcsec\ slit and a 800 grooves/mm grating in first
order were used. The above instrumental set-up gave a spatial scale
along the slit of 0\farcs6 pixel$^{-1}$, a scale perpendicular to the slit
of 0.75\AA\ pixel$^{-1}$, a spectral range of 3200--5200\AA\ and a spectral
resolution of $\sim$ 3\AA\ (FWHM). 
While the signal-to-noise ratio of those spectra is sufficiently 
high to allow for a search of  
low-intensity broad components of strong emission lines,  
no special criterion has been applied in assembling the sample 
to select out only those BCDs that possess broad emission lines. 
Therefore, the MMT subsample of BCDs with no detected broad component 
constitutes a good unbiased sample with which to 
compare the BCDs with broad emission.   

\begin{figure*}
\figurenum{1}
\epsscale{0.9}
\plotone{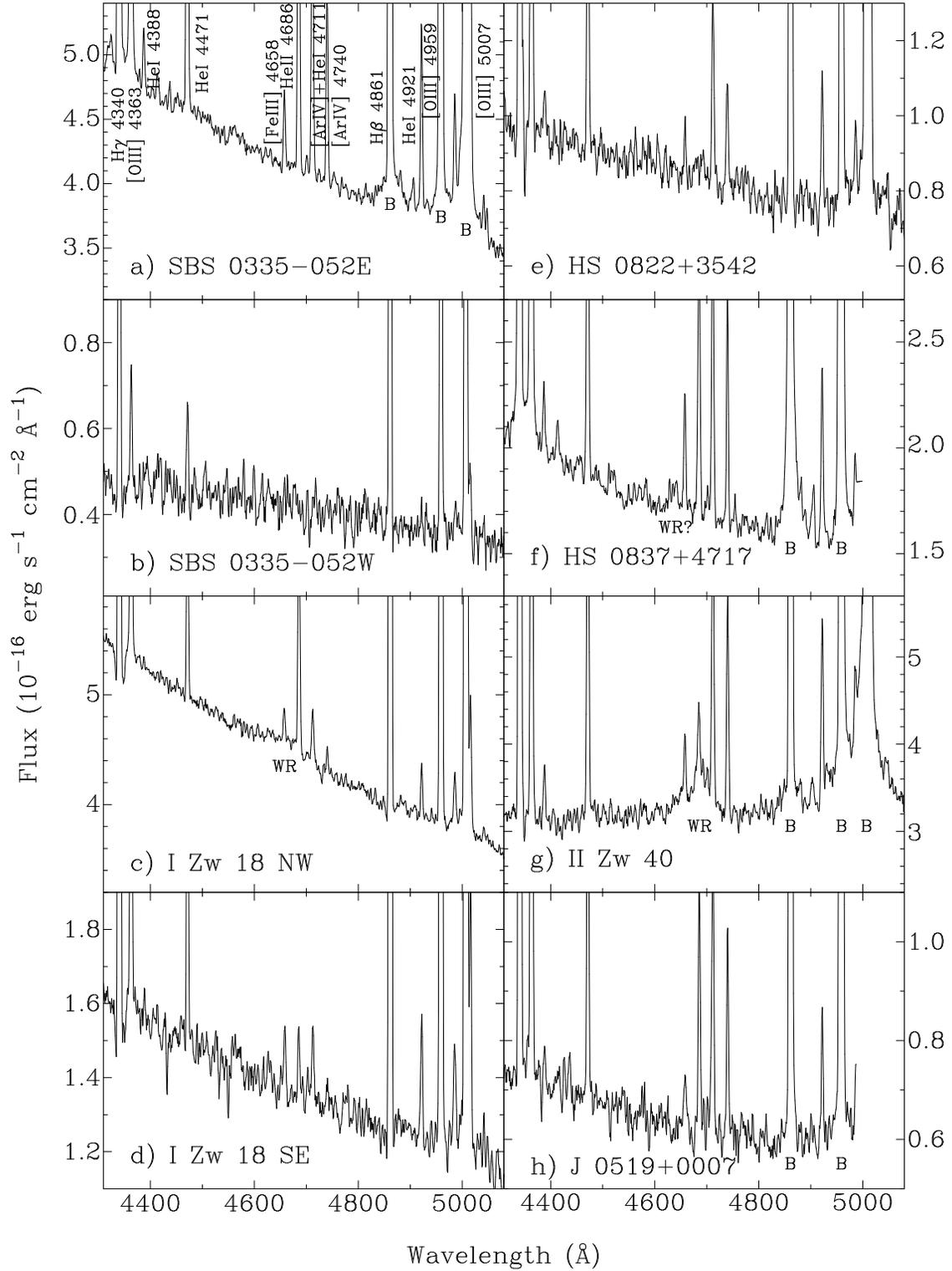}
\figcaption{Redshift-corrected MMT spectra of blue compact dwarf galaxies. The 
narrow emission lines
are labeled in panel a), the WR bump is labeled ``WR'' and the broad components
of strong lines are labeled ``B''. \label{figMMT}}
\end{figure*}

The 23 MMT spectra cover a wavelength region which includes the
blue WR bump at $\lambda$4650 and the strong H$\beta$ and [O {\sc iii}]
$\lambda$4959, 5007 emission lines. They 
are shown in Fig. \ref{figMMT}. The nebular
emission lines are labeled only in Fig. \ref{figMMT}a, 
while the broad components
of strong emission lines and the WR broad lines are respectively 
labeled ``B'' and ``WR'' in all spectra, whenever they are present. 
In total, 14 spectra show broad emission in the MMT sample.
The parameters of the narrow (nar) and broad (br) components 
of strong lines have been obtained by Gaussian fitting 
using the IRAF
SPLOT routine. They are given in Table \ref{tab1}. The fluxes have been 
corrected
for interstellar extinction by adopting 
the extinction coefficient $C$(H$\beta$)
derived by \citet{TI05} from the nebular narrow 
emission lines. The luminosities of
the broad line emission are then obtained from the 
extinction-corrected fluxes. To convert fluxes into luminosities, 
we generally use redshift distances corrected for Virgocentric flow 
with a Hubble constant 
$H_0$ = 75 km s$^{-1}$ Mpc$^{-1}$. Exceptions are the nearby  
objects Mrk 71 and Mrk 209 and the  
H {\sc ii} region J1404+5423 in the spiral galaxy M 101, for which 
we have used  
distances determined by other methods: 
$D$ = 3.42 Mpc for Mrk 71 \citep{TI05b}, $D$ = 5.7 Mpc 
for Mrk 209 \citep{K02} 
and $D$ = 6.7 Mpc for M 101 \citep{Pi04}.

\subsection{The SDSS sample}

In addition to the MMT spectra, we have also searched the 
SDSS DR5 data base for emission-line galaxy spectra showing broad emission.
We first select only those galaxy spectra which possess 
strong emission lines and a detected [O {\sc iii}] $\lambda$4363 emission 
line \citep{I06a}.
This selection criterion presents several advantages. First, 
because 
the [O {\sc iii}] $\lambda$4363 emission line is generally weak -- the  
typical flux is only a 
few percent of H$\beta$ or less -- the spectra of the 
selected galaxies have necessarily a relatively high 
signal-to-noise ratio. This high ratio is  
needed to search for broad line emission as its intensity is 
generally low.  Second, the detection of the 
[O {\sc iii}] $\lambda$4363 emission line 
allows the direct determination of element abundances. 
Excluding obvious high-metallicity AGN by
visual examination of their spectra, we end up with a sample of $\sim$ 10,000
emission-line galaxies. To check that 
our selection criteria work and pick out mostly star-forming galaxies (SFG), 
we have plotted with dots the whole sample in the   
[O {\sc iii}] $\lambda$5007/H$\beta$ -- [N {\sc ii}] $\lambda$6583/H$\alpha$
diagnostic diagram (Fig. \ref{diagnDR5}). The figure  
shows that the vast majority of the points do lie in the 
SFG part of the diagram. 

We then carefully
examine all $\sim$ 10,000 spectra to search for broad line emission.
We have found a total of 21 spectra with evident 
broad line components. These include 19
spectra in 17 BCDs with redshifts in the range 
from 0.004 to 0.3, and two H {\sc ii} regions, J1402+5420 and J1404+5423,
in the spiral galaxy M 101. The general characteristics 
of the SDSS galaxies with broad emission are given in 
Table \ref{tab2}.
The absolute magnitudes of the SDSS objects, excluding the two 
H {\sc ii} regions in M 101,
are in the range $M_g$ $\sim$ --14.8 - --19.3 mag, 
typical of dwarf galaxies and similar to those of the BCDs 
in the MMT sample.
As we will see later, the luminosities of the broad H$\alpha$ lines are in 
general larger in galaxies with a greater broad to narrow component 
flux ratio in H$\alpha$, a ratio which we will denote hereafter by $R$. 
This suggests that broad emission is caused by  
different mechanisms, depending on whether $R$ 
is small or large.
Therefore, we have divided 
the SDSS galaxies in Table \ref{tab2} into two subsamples, 
subsample 1 with $R$ less than 0.25, and subsample 2 with $R$ 
greater than or equal to 0.25. 
Subsample 1 includes 12 spectra. They are shown in the 
whole wavelength range and in the narrow wavelength range around 
H$\alpha$ respectively 
in Figs. \ref{figbr} and \ref{figbr1}. Subsample 2 
includes 9 spectra.  They are shown in the 
same wavelength ranges as for the spectra in subsample 1
 respectively in Figs. \ref{figSN} and \ref{figSN1}.

Like for the galaxies in the MMT sample, 
the fluxes of the narrow and broad components of the relevant 
emission lines were measured using 
Gaussian fitting with the IRAF SPLOT routine. 
The fluxes were corrected for extinction
using the observed decrement of the 
narrow hydrogen Balmer lines. The extinction-corrected fluxes of 
the narrow lines for each galaxy, 
together with the extinction coefficient
$C$(H$\beta$), the equivalent width of the H$\beta$ emission line 
EW(H$\beta$),
the H$\beta$ observed flux $F$(H$\beta$) and equivalent widths of
underlying hydrogen absorption lines EW(abs) are given in Table \ref{tab3} and
Table \ref{tab4} for subsamples 1 and 2 respectively. The physical
conditions and element abundances of the H {\sc ii} regions
in the SDSS sample are derived from narrow line fluxes following \citet{I06a}.
The electron temperatures and number densities, ionic and total element
abundances are shown in Tables \ref{tab5} and \ref{tab6} for subsamples 1
and 2 respectively. 
We find the abundances and the abundance ratios obtained
for the SDSS galaxies to be in the range characteristic of  
low-metallicity dwarf galaxies \citep{I06a}. 
 The parameters of the 
broad components of the relevant 
emission lines are given respectively for subsamples 1 and 2 
in Tables \ref{tab7} and \ref{tab8}.

To study possible time variations of the
broad emission lines, we have paid special attention to 
objects in our sample that 
possess repeated SDSS observations.
Two objects in subsample 1 have each been observed twice during the course
of the SDSS. The first object, J0840+4707 $\equiv$ HS 0837+4717,  
has spectra 0549-621 and 0550-092 obtained respectively 
on 13 March 2001 and on 19 February 2001. 
It was also observed with the MMT in 
1997 \citep{P04} and in 2004 \citep{TI05}. The latter observation 
is included in the MMT sample.
In all these spectra of J0840+4707, broad emission is visible in 
the strong emission lines. The 
second object, J2250+0000, has  
SDSS spectra 0675-039 and 
0676-192, obtained respectively on 10 November 2002 and on 26 
September 2001. A third object, J1404+5423, has both SDSS (24 March 2004) 
and MMT (4 February 2005)
observations \citep{TI05} and is therefore present in both samples.

\section{MECHANISMS FOR  BROAD LINE EMISSION}

Examination of Tables \ref{tab1}, \ref{tab7} and \ref{tab8} shows that 
the range of broad line luminosities in the galaxies 
in both samples is large.
The lowest luminosities are consistent with the emission line 
luminosities of a single or a few massive stars with stellar winds, while the
highest luminosities are typical of bright type II SNe and
  low-luminosity AGN. 
We discuss below different mechanisms which may be responsible for the broad 
emission, by considering in turn the MMT and the SDSS samples.

\subsection{The MMT sample}

\subsubsection{General characteristics}

All H {\sc ii} regions in the MMT sample show strong narrow nebular
emission lines in their spectra, 
suggesting the presence of massive main sequence stars 
(Fig. \ref{figMMT}). However, not all spectra show
broad components in their strong emission lines. We emphasize that these broad
features are real and  
not the result of instrumental effects, as shown by the comparison of 
the H$\beta$ profiles of H {\sc ii} regions in different 
galaxies. For example, the H$\beta$ emission line
in I Zw 18 NW and Mrk 178-1 is narrow, without 
any sign of a broad component, while broad components are evidently 
present in other galaxies 
with a comparable or weaker narrow H$\beta$ emission 
line, such as SBS 0335--052E and HS 0837+4717.
 Furthermore, we have checked the profiles of the strongest lines in the  
comparison lamp spectra obtained with large exposure times before each observing
night. These lines have similar widths and comparable or larger fluxes 
as compared to the strongest lines with detected broad emission in the spectra of
the studied galaxies. However, no broad wings were seen, ruling out scattered light 
in the spectrograph as the cause of the broad lines.
The most prominent broad components are seen in the spectra of the 
3 H {\sc ii} 
regions in Mrk 71. This is in part due to the proximity of 
this galaxy: at a distance of 
3.42 Mpc \citep{TI05b}, Mrk 71 is the nearest galaxy 
in both the MMT and SDSS samples. 
Most often, the broad components are seen only in 
the brightest emission lines, H$\beta$ $\lambda$4861 and [O {\sc iii}]
$\lambda$4959, 5007. In the MMT sample, only two objects have broad 
components that are detected
in other emission lines: Mrk 71-1, where the [Ne {\sc iii}] $\lambda$3868 and 
H$\gamma$ $\lambda$4340 lines also show broad components, 
and HS 0837+4717 which shows 
broad components in several emission lines, including 
[O {\sc iii}] $\lambda$4363. 

Generally, the full widths at half maximum ($FWHM$) of the broad
features are $\sim$ 20\AA, corresponding to an expansion velocity of
$\sim$ 1500 km s$^{-1}$. The luminosities of the broad emission lines vary 
by some 4 orders of magnitude, from $\sim$10$^{36}$
erg s$^{-1}$ in Mrk 209 and Mrk 71-3 to $\sim$10$^{40}$
erg s$^{-1}$ in HS 0837+4717. However, the flux ratios of the broad to 
narrow components for the majority of the MMT objects lie  
in the very narrow range of $\sim$ 1 -- 2\%. Exceptions are 
HS 0837+4717 and Mrk 59-2, where that ratio is significantly higher,
being respectively around 5-14\% and 15\%.
We next consider various mechanisms which may be responsible for the 
broad line emission. 

\subsubsection{Wolf-Rayet stars}

There is not a one-to-one correspondence between WR emission 
and broad emission in the strong lines. 
The WR broad emission lines are present in the spectra of 
10 objects: I Zw 18 NW,
II Zw 40, Mrk 35-1, Mrk 59-1, Mrk 71-2, Mrk 94,
Mrk 178-1, Mrk 209, SBS 0926+606A and J1404+5423.
However, not all of these spectra show broad wings in the 
H$\beta$ and [O {\sc iii}]
$\lambda$4959, 5007 emission lines. In particular, broad emission is weak or
absent in the spectra of I Zw 18 NW, Mrk 178-1 and SBS 0926+606A. 
On the other hand, no significant WR emission is present in the spectra of
SBS 0335--052E, HS 0837+4717 and Mrk 71-1, while broad components are 
clearly seen in the strong emission lines. Thus, 
while we cannot rule out that a 
small part of the broad emission originates in the circumstellar 
envelopes of WR stars, this is not the case for all galaxies. That 
H$\beta$ broad emission in the envelopes of WR stars is weak 
is plausible because WR stars have already lost their hydrogen-rich envelopes,
and their fast moving circumstellar envelopes do not emit hydrogen
lines.

\subsubsection{Stellar winds from non-WR stars and supernovae}

 Examination of the spectra of the MMT sample (Fig. \ref{figMMT}) 
leads us to the following general conclusions:

\noindent 1. Broad emission is seen mainly in galaxies that are 
forming stars in massive
compact super-star clusters, for example in
SBS 0335--052E, Mrk 71, Mrk 209, HS 0837+4717. On the other hand, in galaxies 
where star formation occurs in more diffuse environments, like  
in I Zw 18NW, I Zw 18SE,
SBS 0335--052W and SBS 0940+544, broad components are absent. \\
2. Metallicity does not appear to play a role. 
For example, SBS 0335--052E which has broad emission has about the 
same metallicity as I Zw 18 which does not. \\
3. Broad emission is mainly seen in galaxies with high equivalent widths
of H$\beta$, i.e., with younger H {\sc ii} regions. \\  
4. Broad emission is seen over a period of 
at least several years, as repeated observations of several 
objects over that time scale show. It is likely that it exists over even  
longer periods. \\ 
5. In one object, HS 0837+4717, the broad H$\alpha$/H$\beta$ flux ratio
is significantly larger than the recombination value. This suggests
either high extinction in the regions with broad emission or, more likely, 
that the region
of broad emission is very dense ($N_e$ larger than 10$^4$ cm$^{-3}$),
 resulting in collisional enhancement 
of the hydrogen lines. Thus, broad emission in this galaxy is more likely
associated with a dense circumstellar medium than a less dense interstellar
medium. 

All these facts suggest that, for most of the MMT galaxies, the
most likely source of broad H$\beta$ emission is a dense gaseous 
medium associated with young massive stars.  
This dense medium can be either circumstellar
envelopes or stellar winds around Ofp/WN9
or Luminous Blue Variable (LBV) stars.
These stars have masses in the range 
of $\sim$ 30 to 100 $M_\odot$. They lose their outer
hydrogen-rich layers through a wind at a rate 
$\dot{M}$ $\sim$ 10$^{-6}$ -- 10$^{-3}$ $M_\odot$ yr$^{-1}$ 
and with terminal velocities ranging
 from several hundred to $\sim$ 2000 km s$^{-1}$ 
\citep{C75,HA92,LL93,HD94,S94,N96,P96,D97,D01}. This corresponds to the  
range of velocities deduced from the broad emission lines.

SNe can also play a role as sources of 
broad emission. 
 Thus, broad emission can be produced by type IIn
SNe (a description of the different SN types will be given in 
\S \ref{s3.3}). In this case, it arises from 
shock interaction of the blast wave with the slow-moving circumstellar
envelope of the star expelled just prior to its explosion. 
Type IIp SNe are probably excluded, at least
for some objects, because they emit appreciably only on 
the short time scale of less than a year. In particular, 
for galaxies with spectra taken at different times, 
broad emission is seen 
on time scales of more than a year
[e.g., in HS 0837+4717, this paper and \citet{P04}; in SBS 0335--052E,
this paper and \citet{I06}], comparable to the time scale of some
  IIn SNe \citep[e.g., ][]{F97,P02}.
The type IIn SN mechanism can easily be distinguished from the closely related SN
remnant/superbubble mechanism discussed below (\S\ref{bubble}). In the former case, 
broad emission originates in the dense circumstellar envelope.
Therefore, a steep decrement of the broad Balmer hydrogen lines caused
by collisional excitation is expected. This is not so in the SN
remnant/superbubble case.

\subsubsection{Supernova remnants and supernova bubbles \label{bubble}}

Another likely source of broad emission is remnants produced by single or 
multiple supernova events and propagating in the
interstellar medium, as suggested by \citet{I96}. In this case, the flux ratio
of the broad to narrow components should be 
comparable to the ratio of the mass of the massive  
exploded stars to the mass of gas which is ionized by the UV radiation
from those stars. \citet{I96} estimated this value to be $\sim$ 1\%, 
nearly independently of the total number of massive stars. The fact that, 
despite a large range in $L_{br}$ luminosities,
the $I_{br}$/$I_{nar}$ ratio stays in the relatively narrow range of a few 
percent (Table \ref{tab1}), appears 
to support this mechanism for the 
majority of the galaxies in the MMT sample. 

The multiple supernova bubble phenomenon was discussed in detail 
by \citet{R92} in the case of Mrk 71. Those authors  
 showed that the broad emission is spatially extended, leading 
them to consider a superbubble blowout as the cause. However, as discussed 
in the Introduction,    
there are problems with the superbubble model and the origin 
of the broad emission in Mrk 71 remains unclear. Other galaxies
in the MMT sample are more distant, making it difficult to study the spatial
distribution of the broad emission. There is some evidence that it
is extended in SBS 0335--052E \citep[e.g., ][]{I06}. 
Thus, multiple SNe events and superbubbles associated 
with massive stars are possible in some galaxies, 
but because of the difficulties 
discussed by \citet{R92} with the superbubble model, 
they are not likely to be the main originators of broad emission.
Additionally, such an explanation would not work 
for such an object as HS 0837+4717. 
Its H$\alpha$/H$\beta$ flux ratio for broad emission is $\sim$ 9,  
significantly larger than the value expected from recombination, and
suggesting that high extinction or   
collisional excitation is at work in high-density regions, such as the 
circumstellar envelopes of Of and LBV stars.

\subsubsection{Low-luminosity AGN}

The broad H$\beta$ luminosities in three objects from the MMT
  sample, SBS 0335--052E, HS 0837+4717 and J0519+0007, are 
in the range 4 $\times$ 10$^{38}$ -- 3 $\times$ 10$^{40}$ erg s$^{-1}$, 
 comparable to those of the lowest-luminosity AGN, such as NGC 4395
  \citep{FS89}. Therefore, the accretion
  of gas onto intermediate-mass (10$^3$ -- 10$^5$ M$_\odot$) black holes as a
  mechanism of broad-line emission is a possibility in these objects. 
  At least two of these BCDs, SBS 0335--052E and HS 0837+4717, 
likely contain very compact and dense super-star
  clusters \citep{TI05}, which favors the formation of black holes. 
The metallicities of these BCDs (2-4\% that of the Sun) are
  considerably lower than those of the lowest-luminosity AGN known. 
For example, the oxygen abundance of NGC 4395 is 
12 + log O/H = 8.45 $\pm$ 0.22 \citep{vZ06}, slightly lower than the solar
value. As pointed out before, at the
  low metallicities of the galaxies in our sample, it is difficult to use 
emission-line diagnostic diagrams to distinguish between AGN and
  star-forming regions: in the [O {\sc iii}]
  $\lambda$5007/H$\beta$ vs [N {\sc ii}] $\lambda$6583/H$\alpha$ diagram,
these two classes of objects lie in overlapping regions 
  \citep{Gr06,S06}. However, hard radiation characteristic of AGN is present 
in those objects as evidenced by the relatively strong He {\sc ii}
  $\lambda$4686 emission line (ionization potential of 4 Ryd) 
in the spectra of all these
  galaxies (Fig. \ref{figMMT}a,f,g). Additionally, the [Ne {\sc v}]
  $\lambda$3426 emission line (ionization potential of 7 Ryd) 
is detected in the spectra of SBS 0335--052E
  and HS 0837+4717 \citep{TI05}, although it is not evident in the
  spectrum of J0519+0007 \citep{IT07}.

There are nevertheless several problems with explaining broad hydrogen
lines by accretion of gas onto an intermediate-mass 
black hole:  1) it is not clear whether 
there is enough time for the formation of such black holes: the stellar 
clusters
in all three galaxies are very young, with an age of $\sim$ 3 -- 4
Myr, as deduced by their H$\beta$ equivalent widths; 
2) numerous O-stars in compact clusters can sweep out the gas from
the cluster, precluding the feeding of a black hole; 3) the 
He {\sc ii} $\lambda$4686 line emission is extended, at least in  
the best studied case, the BCD SBS 0335--052E \citep{I01,I06}, 
which favors shocks instead of a compact object as its origin.

We conclude that, for most of the MMT sample galaxies with broad H$\beta$
line luminosities in the range 10$^{36}$ - 10$^{40}$ erg s$^{-1}$, the
most likely source of broad emission is stellar winds from Ofp/WN9
or LBV stars.
In the low luminosity range of  
10$^{36}$ -- 10$^{37}$ erg s$^{-1}$, the broad H$\beta$ line emission can be  
explained by the presence of only one or a few Ofp/WN9 or LBV stars
\citep[e.g.][]{D97}. These galaxies and H {\sc ii} regions are J1404+5423,
Haro 3, Mrk 209 and Mrk 71. Their spectra show broad
WR features and, in principle, part of the 
broad features can also be due to winds from WR
stars. However, the H$\beta$ broad component 
luminosities are significantly 
larger in SBS 0335--052E,
HS 0837+4717 and J0519+0007. These galaxies do not appear to contain
WR stars, so we cannot invoke these stars to explain the broad emission. 
If stellar winds from Ofp/WN9 and LBV are responsible for the broad 
component, then the number of these stars in those three galaxies 
should be more than $\sim$ 100. In those objects with the 
highest broad H$\beta$ line luminosities (10$^{39}$ - 10$^{40}$ erg s$^{-1}$),
two other mechanisms can also be considered as likely sources of broad emission: 
fast shocks produced by multiple SN events propagating in the 
interstellar medium \citep{TI05} and/or shocks propagating in the 
circumstellar envelopes of type IIn SNe.  

\subsection{The SDSS subsample 1 with $R$ $\leq$ 0.25}

\subsubsection{General characteristics}

We next examine the different mechanisms that may be responsible for 
broad emission in the SDSS sample. We will discuss in turn subsample 1 
which contains galaxies with a H$\alpha$ flux ratio of the 
broad component to the narrow component 
less than 0.25 and subsample 2 which contains galaxies with that ratio larger 
than 0.25.

The parameters of the emission lines with broad components 
are shown for subsample 1 
in Table \ref{tab7}. H$\alpha$ broad components are seen in
all 12 spectra (Fig. \ref{figbr}, \ref{figbr1}). Their $FWHM$s of
$\sim$ 20 -- 40\AA\ correspond to velocities 500 -- 1000
km s$^{-1}$, similar to those found for the galaxies with broad lines in the
MMT sample. However, the flux ratios of the broad to narrow components of the 
galaxies in the
SDSS subsample 1 are higher than those 
of the galaxies in the MMT sample. This
difference is most likely due to a selection effect. SDSS 
spectra have generally a lower signal-to-noise ratio than the MMT spectra.
Therefore, broad emission can only be detected in 
galaxies with relatively strong broad lines.
The galaxy HS 0837+4717 which is present in both the MMT and SDSS
(J0840+4707, spectra 0549-621 and 0550-092) samples, has indeed  
the largest luminosities of the broad H$\alpha$ and/or H$\beta$ emission lines
in each sample.
We note also, that different spectra of HS 0837+4717, obtained at different
epochs over a time span of several years, show similar strengths of 
the H$\alpha$ and H$\beta$ lines (compare Tables \ref{tab1} and \ref{tab7}).

\subsubsection{Circumstellar envelopes around O and LBV stars and 
supernova remnants}

In two objects, J0840+4707 (spectra 0549-621 and 0550-092) 
and J1404+5423 (spectrum 1324-234), broad emission is present 
in the H$\beta$ line. The H$\alpha$
to H$\beta$ flux ratio of the broad components in both spectra is significantly
larger than the recombination value, suggesting that the broad 
hydrogen lines are formed not in the H {\sc ii} region, but in the denser
circumstellar envelopes. There are several other lines of evidence which  
suggest that broad 
emission originates from dense regions. In the spectrum of J1404+5423,
several lines show broad components. The luminosities of these broad lines 
range from $\sim$ 10$^{36}$ to $\sim$ 10$^{37}$ erg s$^{-1}$, 
close to those expected for  
the line luminosities of a single massive star with a stellar wind
\citep[e.g. ][]{D97}. The high density
and consequently the 
high optical depth in some permitted lines in the spectrum of J1404+5423 is 
indicated by the following facts (Table \ref{tab7}): 1) a large 
broad H$\alpha$/H$\beta$ flux 
ratio; 2) the 
[O {\sc iii}] $\lambda$4363/($\lambda$4959+$\lambda$5007) broad line ratio
is significantly larger than the one 
expected for a H {\sc ii} region, even for a very hot one 
with $T_e$ $\sim$ 20000K. This large ratio is most likely due
to a reduction of the fluxes of the broad $\lambda$4959 and $\lambda$5007 
lines by collisional de-excitation, as compared to the low-density
case. A similar effect has been discussed by \citet{T96} for the BCD Mrk 996; 
3) the line intensities relative to H$\beta$ 
of $\sim$ 0.23 and 0.15 respectively for the 
broad $\lambda$5876 and especially the $\lambda$7065 He {\sc i}
emission lines are signicantly 
larger than those seen in typical H {\sc ii} regions ($\sim$ 0.1 and 0.02). The
high relative fluxes of the He {\sc i} lines are likely caused by significant
collisional excitation from the metastable 2$^3$S level; 4) a very strong broad
auroral [N {\sc ii}] $\lambda$5755 emission line is unusual for a normal
H {\sc ii} region but is often associated with the nitrogen-enriched
regions around LBV stars. 

Thus, the spectrum 1324-234 of J1404+5423 most likely resembles that 
of a high-excitation H {\sc ii} region superimposed on the spectrum of a 
Ofp/WN9 or LBV star. The broad $\lambda$4650 and $\lambda$5808 emission lines 
characteristic of WR stars are not seen. 
However, WR features are seen in the MMT spectrum
of this object obtained on 2005 February 4 (Fig. \ref{figMMT}y), 10 months
after the SDSS spectrum. There are other differences between
the MMT and SDSS spectra of J1404+5423. The broad H$\beta$, [O {\sc iii}]
$\lambda$4959, $\lambda$5007 luminosities in the SDSS spectrum are more than 
two times larger than those in the MMT spectrum. These luminosity 
differences cannot
be understood by differences in apertures since the aperture used
in the SDSS observations has an area not larger but
smaller (by a factor of two) than the one used in the MMT
observations. Instead, the luminosity differences can be as explained 
by one or both of the following possibilities:  
1) the LBV has undergone an outburst at the beginning
of 2004 which quickly faded after one year; 2) because of pointing 
errors, slightly different
regions were sampled during the SDSS and MMT observations.

A similar conclusion can probably be 
drawn for the spectra 0549-621 and 0550-092
of the galaxy J0840+4707. However, the luminosity
of the broad H$\alpha$ line in this galaxy is some 
two orders of magnitude larger than the one 
in J1404+5423, and thus several hundreds of Ofp/WN9 or LBV stars 
are needed to account for such a large broad line luminosity.
SNe remnants may also contribute to the broad luminosity 
of J0840+4707 \citep{P04}. 
There is not enough information for other galaxies in subsample 1 
to constrain the mechanism of broad line emission, 
as only a broad H$\alpha$ line is seen (Table \ref{tab7}, Figs. \ref{figbr},
\ref{figbr1}). However, the similarity in such properties as the widths 
and the luminosities of the broad 
H$\alpha$ lines in the subsample 1 objects suggests that stellar winds
from Ofp/WN9 or LBV stars and, to a lesser part supernova remnants, 
provide the most likely mechanism for the formation of broad emission lines. 
Finally, 
we note that the highest broad H$\alpha$ luminosities of the galaxies 
in subsample 1 (10$^{40}$ -- 10$^{41}$ erg s$^{-1}$) compare well with 
those of low-luminosity AGN. 
Therefore, we cannot exclude the
presence of AGN in these galaxies.  However, the lack of hard 
radiation as evidenced by the weak
He {\sc ii} $\lambda$4686 and 
[S {\sc ii}] $\lambda$6717, 6731 emission lines, would appear to 
argue against such a possibility. Low-luminosity 
AGN will be discussed more in detail in 
\S\ref{s4}.

\subsection{The SDSS subsample 2 with $R$ $>$ 0.25 : SN or AGN as 
possible sources of broad emission \label{s3.3}} 

The galaxies in subsample 2 possess 
the highest ratio of broad to narrow H$\alpha$ emission line
fluxes in the SDSS sample (Table \ref{tab8}). 
Their broad components show a variety of 
properties as can be seen in Figs. \ref{figSN}, \ref{figSN1}. They 
are most likely produced in type II
SNe or/and AGN. Three major classes of type II SNe are often 
considered, based on their luminosity, the 
shape of their broad H$\alpha$ 
emission line and the time variation of their total brightness
\citep{F97,T07}: 
1) type IIp (where p stands for ``plateau'') with  
a plateau in brightness during $\sim$ 100 days after maximum.
Their H$\alpha$ emission lines are broad and show P Cygni profiles during the 
first $\sim$ 200 days; 2) type IIl (where l stands for ``linear'') 
characterized by a rapid, steady linear
decline of the brightness in the same period.
 They therefore become faint on timescales of 
$\la$ 100 days. Their H$\alpha$ is broad but no P Cygni profile is observed;
3) type IIn (where n stands for ``narrow'') characterized by a 
much slower decline of
the brightness as compared to type IIp SNe. In contrast to spectra of types 
IIp and IIl SNe,
broad absorption lines are weak or absent in the spectra of type IIn SNe. 
Instead, these 
are dominated by strong narrow H$\alpha$ and other emission lines, 
believed to be the sign of energetic 
interaction between the SN ejecta and dense 
circumstellar gas. The H$\alpha$ luminosities of type IIn SNe are
 larger than those of type IIp and IIl SNe. Alternatively, 
broad emission can also be due to gas accretion onto a black hole, as 
the highest broad line H$\alpha$ luminosities of the galaxies in subsample 2 
are in the range of low-luminosity AGN (10$^{40}$--10$^{42}$ erg s$^{-1}$). 

Since we have no data on the temporal evolution
of the brightness of our objects, we classify them 
mainly on the basis of their luminosities and the shape of 
their H$\alpha$ emission line. Because type IIl SNe decrease quickly 
in brightness 
and become faint after $\sim$ 100 days, it is unlikely that the SDSS 
contains spectra of this type. We will make use only of the categories IIp and 
IIn to classify SN spectra. 
Since these objects are very rare - we find only 9 such objects
in the SDSS DR5 -- we describe and classify each galaxy's spectrum separately.

{\it J2230$-$0006 (spectrum 0376-176)}. Several hydrogen lines in this object 
show asymmetric profiles
with $FWHM$s $\sim$ 12 -- 23\AA. All hydrogen lines show blueshifted (by
$\sim$ 800 km s$^{-1}$) absorption. Additionally, the Balmer decrement for the 
broad hydrogen lines is considerably larger
 than the recombination value. This 
suggests that broad emission originates in a very dense gas. 
The luminosities
of the broad hydrogen lines can be understood in terms of a single IIn SN. 
The temporal evolution of emission lines showing a similar velocity
structure has been discussed e.g. by \citet{F97}, \citet{C91}, 
\citet{F01}, \citet{SS91} and \citet{C04} for different type IIn SNe. 
Alternatively, the luminosities of the broad lines can be
explained by one or several extremely luminous LBV stars, comparable to or 
brighter than $\eta$ Car.
The continuum is blue and is typical
of BCDs. The extinction coefficient derived from the Balmer decrement of the 
narrow hydrogen lines, $C$(H$\beta$) = 0.245 (Table \ref{tab4}) is also 
in the range obtained for BCDs. The spectrum shows no 
strong He {\sc ii} $\lambda$4686,
[O {\sc i}] $\lambda$6300 and [S {\sc ii}] $\lambda$6717, 6731 emission lines,
making the presence of intense hard radiation produced by a high-mass AGN 
unlikely. However, the broad hydrogen lines in this galaxy 
can in principle be produced by gas accretion onto an  
intermediate-mass black
hole. Therefore, the presence of a low-luminosity AGN cannot be excluded.

{\it J0045$+$1339 (spectrum 0419-137)}. Only the H$\alpha$ emission line shows 
a broad symmetric component. Its Full Width at Zero Intensity 
($FWZI$) corresponds to a velocity of 
$\sim$ 3000 km s$^{-1}$. 
Instead of rising in the blue, the continuum is flat, which is atypical of BCDs. 
The flatness of the spectrum cannot be explained by
dust extinction as the extinction coefficient is small, $C$ = 0.155. 
It is also likely not due to the light contribution of an old stellar
population in the galaxy as no strong stellar absorption line, such as 
K Ca {\sc ii} $\lambda$3933, is seen in the spectrum.
Rather, it can be caused by a large contribution of
a red continuum that is common for type II SNe at $\ga$ 30 -- 40 days 
after the SN explosion \citep{F97}. 
A weak [Ne {\sc v}] $\lambda$3426 line 
is possibly present, with an intensity of
$\sim$ 4\% that of H$\beta$, but He {\sc ii} $\lambda$4686 is absent. 
However, the quality of the spectrum is not
good enough to make firm conclusions on the presence of hard
radiation. The H$\alpha$ luminosity of the broad 
component is  
2.74$\times$10$^{41}$ erg s$^{-1}$. No blueshifted broad absorption is seen. 
If produced by a type IIn SN, this luminosity is 
among the highest ever observed for this type of SNe 
\citep{CD94,H97,F97,A99}. However, a low-luminosity 
AGN can also account for this broad line luminosity.

{\it J0052$-$0041 (spectrum 0692-187)}. The broad 
component is only present in the H$\alpha$ emission line, with
a $FWZI$ corresponding to an expansion velocity of $\sim$ 3000 km s$^{-1}$.
No spectral signature characteristic of a non-thermal source of 
hard radiation is present. The luminosity of the broad component in the 
H$\alpha$ line is in the range
observed for type IIp SNe \citep{F97}, but it also is comparable
  to that of a low-luminosity AGN.
Most likely, the spectrum is that of a type IIp SN, superposed on that of a 
H {\sc ii} region
with a blue continuum and narrow emission lines. The extinction coefficient 
$C$ = 0.190 is typical of BCDs.

{\it J1047$+$0739 (spectrum 1001-363)}. The Balmer 
hydrogen H$\gamma$, H$\beta$ and H$\alpha$ lines, and the 
He {\sc i} $\lambda$5876, $\lambda$7065 lines all show broad components. 
The spectrum also shows strong nebular lines of the high-excitation H {\sc ii}
region. Weak broad N {\sc iii} $\lambda$4640, 
C {\sc iv} $\lambda$4658 and He {\sc ii} $\lambda$4686 lines suggest
the presence of late nitrogen and early carbon Wolf-Rayet stars.
The continuum is flat, which can be explained in part
by a relatively large extinction coefficient $C$(H$\beta$) = 0.355. 
However, we note
that the value of $C$(H$\beta$) in this object is not well determined 
because of uncertainties
in the extraction of the narrow hydrogen lines from the total line profiles,
especially for the H$\alpha$ line. 
The flat continuum cannot probably be explained by the light contribution of
an old stellar population in the galaxy because no strong stellar absorption 
line is seen in the spectrum.
The $FWZI$ of the broad H$\alpha$ line 
corresponds to an expansion 
velocity of $\sim$ 2000 km s$^{-1}$. The Balmer decrement for the broad
hydrogen lines is very large, suggesting collisional excitation in 
high-density regions. Not only the broad, but also the narrow He {\sc i}
$\lambda$5876 and $\lambda$7065 emission lines are very strong compared to 
H$\beta$, while the He {\sc i} 3889 line is weak. This suggests that both the  
broad line and narrow line emission originate in regions which are optically 
thick in the He {\sc i} 3889 line. \citet{Z06} have classified this spectrum as
that of a narrow-line Seyfert 1 galaxy. They note that J1047$+$0739
is the only one in their sample of $\sim$ 2000 narrow-line Seyfert 1 
galaxies with [O {\sc iii}] $\lambda$5007/H$\beta$ $>$ 3 and 
with the width of a broad 
H$\alpha$ and H$\beta$ less than 2000 km s$^{-1}$. 
However, this spectrum can also be classified as that of a type IIn SN.
\citet{F89} and \citet{H97} have pointed out that in some cases spectra
of SNe are quite similar to those of Seyfert 1 galaxies. 
To support our contention, we do not find any signature of 
hard nonthermal radiation in the spectrum. In particular, 
the [Ne {\sc v}] $\lambda$3426 and He {\sc ii} $\lambda$4686 emission lines 
are absent. If the 
spectrum is that of a type IIn SN, then the H$\alpha$ luminosity of the 
broad component, equal to  
1.57$\times$10$^{42}$ erg s$^{-1}$ (Table \ref{tab8}), 
is the highest one observed thus far in this  
type of SNe, even exceeding the luminosity of the 
``Seyfert 1'' type IIn SN 1997ab 
\citep{H97} and 1988Z \citep{A99}.

{\it J1144$+$5355 (spectrum 1015-019)}. The blueshifted absorption
near the H$\beta$ line and the 
very broad H$\alpha$ line with a P Cygni profile and a 
$FWZI$ corresponding to an expansion velocity of $\sim$ 7000 km s$^{-1}$ are 
suggestive of a type IIp SN, observed at $\sim$ 100 days after the explosion. 
The broad continuum depression shortward of H$\gamma$ is probably due to 
the blueshifted H$\gamma$ absorption line in the spectrum of the SN.
The luminosity of 3.24 $\times$ 10$^{40}$ ergs s$^{-1}$
of the broad H$\alpha$ line is in the range of luminosities usually 
observed for this type of SNe.

{\it J0824$+$2954 (spectrum 1207-512)}. A broad H$\alpha$ line 
with a P Cygni profile 
and a $FWZI$ corresponding to an expansion velocity of $\sim$ 3000 
km s$^{-1}$,
 and broad Na {\sc i} $\lambda$5889, 5895 lines with a P Cygni profile 
are seen.  Broad absorption
is likely present in the Fe {\sc ii} $\lambda$4924, 5018, 5169 and Ca {\sc ii} 
$\lambda$8498, 8542, 8662 lines. The luminosity of
the broad H$\alpha$ line is characteristic of a type IIp SN.
All this 
suggest that we have here the spectrum of a type IIp 
SN, caught a few hundred days after the explosion. 

{\it J1644$+$2734 (spectrum 1690-360)}. The presence of the broad  
[O {\sc i}] $\lambda$6300, 6363 lines, the 
H$\alpha$ line which also has a P Cygni profile, and the 
[Ca {\sc ii}] $\lambda$7300 and Ca {\sc ii} 
$\lambda$8498, 8542, 8662 lines suggest that this is the 
spectrum of a type IIp SN in the nebular stage, several hundred days after 
the explosion. The $FWZI$ of the broad H$\alpha$ line corresponds
to an expansion velocity of $\sim$ 3000 km s$^{-1}$ and its luminosity
is in the range typical for this type of SNe. J1644$+$2734 
is the only
object in our sample that is included in the SN list posted on line by the 
SDSS Spectroscopic Supernova Search group (http://cheops1.uchicago.edu/pub/).

{\it J1025$+$1402 (spectrum 1747-337)}. A strong broad H$\alpha$ line, with
a $FWZI$ corresponding to an expansion velocity of 
$\sim$ 3000 km s$^{-1}$, is present. Its profile shows an absorption feature 
which is blueshifted by $\sim$ 200 km s$^{-1}$, suggesting a high
optical depth. However, the broad component of the H$\beta$ line is weak 
(Fig. \ref{figSN}h). This results in a high 
H$\alpha$ to H$\beta$ flux ratio for the broad components, suggesting
that the broad hydrogen lines are due to collisional excitation 
  in a high-density region. A high density is also indicated by the
  low value of the [S {\sc ii}] $\lambda$6717/$\lambda$6731 line
  ratio. A weak broad N {\sc iii} $\lambda$4640 emission line is
  present, suggesting the presence of late nitrogen Wolf-Rayet stars.
The continuum is flat, suggesting high dust
extinction. This hypothesis is supported by the presence of strong narrow 
Na {\sc i} $\lambda$5889, 5895 interstellar absorption lines.
However, the extinction coefficient $C$(H$\beta$) = 0.015, derived from the  
decrement of the narrow hydrogen Balmer lines, is small. 
Apparently, the regions of broad and narrow emission are 
spatially distinct. 
It is not likely that the reddening is caused by a significant light
contribution of the galaxy's old stellar population because no strong stellar
absorption feature, such as the K Ca {\sc ii} line or the 
TiO bands, is seen in the spectrum.
This galaxy is the one
with the lowest oxygen abundance, 12 + log O/H = 7.36 $\pm$ 0.08, in both 
the SDSS subsamples 1 and 2. The luminosity
of the broad H$\alpha$ line, equal to 3.21$\times$10$^{41}$ erg s$^{-1}$,
 is among
the largest in our sample. A weak He {\sc ii} $\lambda$4686
  emission line with an intensity of 3.8\% that of H$\beta$ is
  present, suggesting the presence of hard ionizing radiation. 
However, the [O {\sc i}]
  $\lambda$6300 and [S {\sc ii}] $\lambda$6717, 6731 emission lines
  are very weak. Therefore, it is not clear whether a source of non-thermal
 radiation is present in this galaxy. We conclude that this is likely
the spectrum of a type IIn SN, superposed on the high-excitation spectrum of a 
H {\sc ii} region. The red continuum can be explained by the SN contributing
 a relatively large fraction of  
the total light of the
  galaxy, similar to the case of J0045+1339. However, the presence of an 
AGN cannot be excluded.

{\it J1222$+$3602 (spectrum 2003-167)}. Several hydrogen and forbidden lines 
show broad components in this spectrum.
 The $FWZI$ of the broad H$\alpha$ line corresponds to an expansion velocity 
of $\sim$ 2000 km s$^{-1}$. Its luminosity of 
2.80$\times$10$^{41}$ erg s$^{-1}$ is the fourth largest one in our
sample. The high H$\alpha$/H$\beta$ and 
[O {\sc iii}] $\lambda$4363/$\lambda$5007 flux ratios indicate a high density
for the emitting region(s). The nebular spectrum in the galaxy is that of a 
very high excitation star-forming H {\sc ii} region. However, it can also 
be that of the NLR of an AGN.  
A strong narrow [O {\sc iii}] $\lambda$4363 line is seen on the
top of the broad line,
with a flux equal to $\sim$ 24\% that of H$\beta$. 
However, the continuum is not blue as expected from a H {\sc ii} region, but 
 flat. This flatness 
cannot be explained by dust extinction since the extinction coefficient
derived from the Balmer decrement of the narrow hydrogen lines is 
small, $C$(H$\beta$) = 0.070. It cannot also be explained by the 
light contribution of the galaxy's red stellar population as no strong stellar 
absorption feature is seen. 
There is no strong signature for the presence of intense hard
non-thermal radiation: the He {\sc ii} $\lambda$4686 emission
line is weak, $\sim$ 2\% of H$\beta$ and the [Ne {\sc v}] $\lambda$3426
emission line is absent.  
It is likely, therefore, that 
the broad features are produced by a type IIn SN. The presence of an 
AGN cannot however be ruled out: the very
  strong [O {\sc iii}] $\lambda$4959, 5007 and [Ne {\sc iii}]
  $\lambda$3868 emission lines are more typical of the NLR of an AGN, than
  of an H {\sc ii} region ionized by massive stars.

\section{CAN SOME OF THE BROAD LINE GALAXIES HARBOR LOW-LUMINOSITY 
AGN? \label{s4}}

Our analysis of the broad emission line properties in the spectra of BCD 
galaxies in both the MMT and the SDSS samples 
shows that they possess a variety of properties. However, they do have a 
common feature: their broad emission is always somehow 
related to the evolution of
massive stars. In many cases, the relative hydrogen line intensities 
suggest a very dense emitting medium, with $N_e$ $\ga$ 10$^4$ cm$^{-3}$. 
In some galaxies, the broad features are caused by individual
SN events. In others, stellar winds and perhaps 
multiple SN events are responsible. 
While the presence of type IIp SN in the spectra of 4 galaxies in 
subsample 2 (J0052--041, J0824+2954, J1144+5355 and J1644+2734) 
is reasonably certain,
it is not clear from the available data whether the 
broad lines in the remaining galaxies of 
subsample 2 are produced by SN events or by some other mechanism.
For example, can they be produced by AGN?
We do not find strong evidence for AGN ionization in the spectra which we
classify as belonging to type IIn SN/AGN.
In these spectra, 
there is no evident spectroscopic sign for the presence 
of an intense source of hard non-thermal radiation: there are no strong 
[Ne {\sc v}] $\lambda$3426, [O {\sc ii}] $\lambda$3727,
He {\sc ii} $\lambda$4686, [O {\sc i}] $\lambda$6300, 
[N {\sc ii}] $\lambda$6583 and 
[S {\sc ii}] $\lambda$6717, 6731 emission lines, as usually found in the 
spectra of AGN. Note however that the weakness of the
 [O {\sc i}] $\lambda$6300 and
[N {\sc ii}] $\lambda$6583 emission lines may be explained, not by 
the absence of an AGN, but by the low
metallicities of our galaxies.
All the type IIn SN/AGN objects were not detected in the NRAO VLA Sky Survey (NVSS) 
suggesting that they are faint radio sources, with 1.4 GHz fluxes less than 1 mJy.

\begin{figure*}
\figurenum{2}
\epsscale{0.9}
\plotfiddle{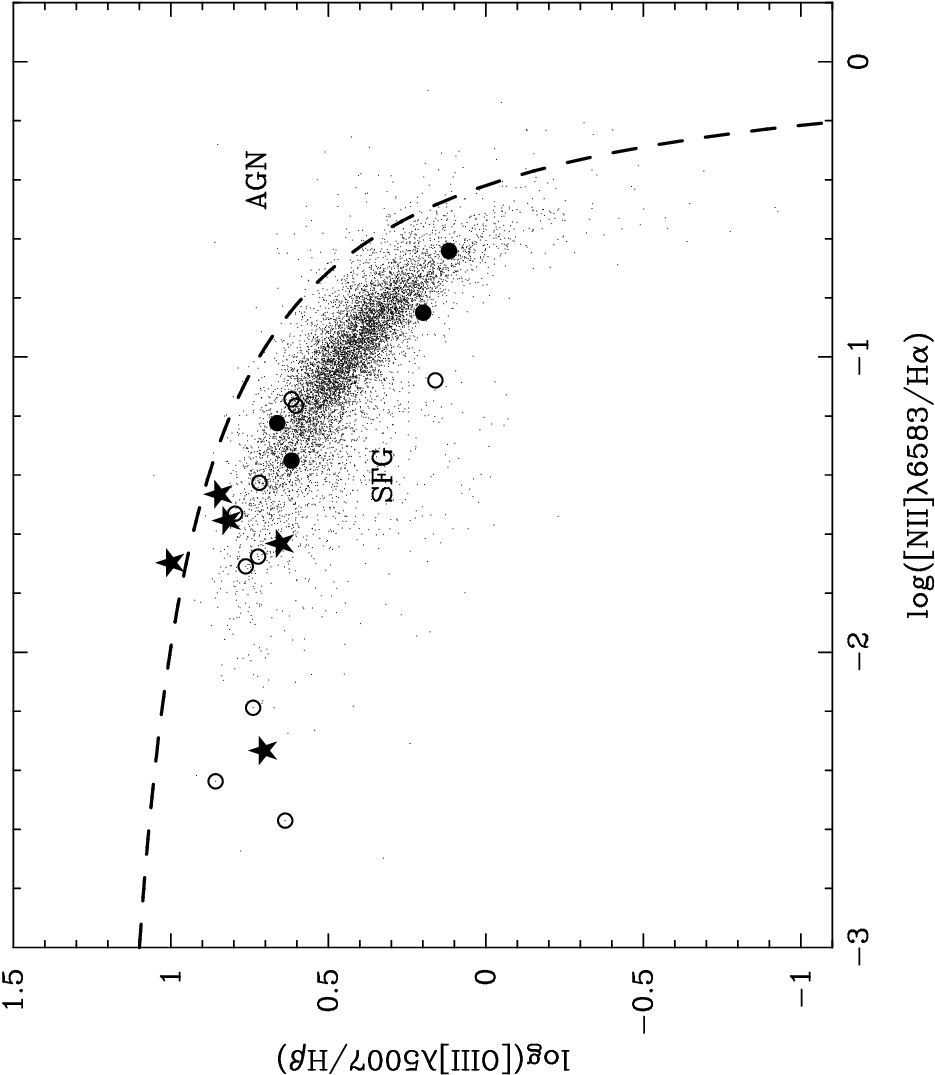}{1pt}{-90.}{420.}{450.}{0.}{0.}
\figcaption{Diagnostic diagram for SDSS galaxies. 
The dashed line separating star-forming galaxies (SFG) 
from AGN is from \citet{K03}. Stars and filled circles represent galaxies
from SDSS subsample 2 with type IIn SN/AGN spectra and with type IIp
SN spectra, respectively. Open circles
are galaxies from SDSS subsample 1. \label{diagnDR5}}
\end{figure*}

\begin{figure*}
\figurenum{3}
\epsscale{1.1}
\plotfiddle{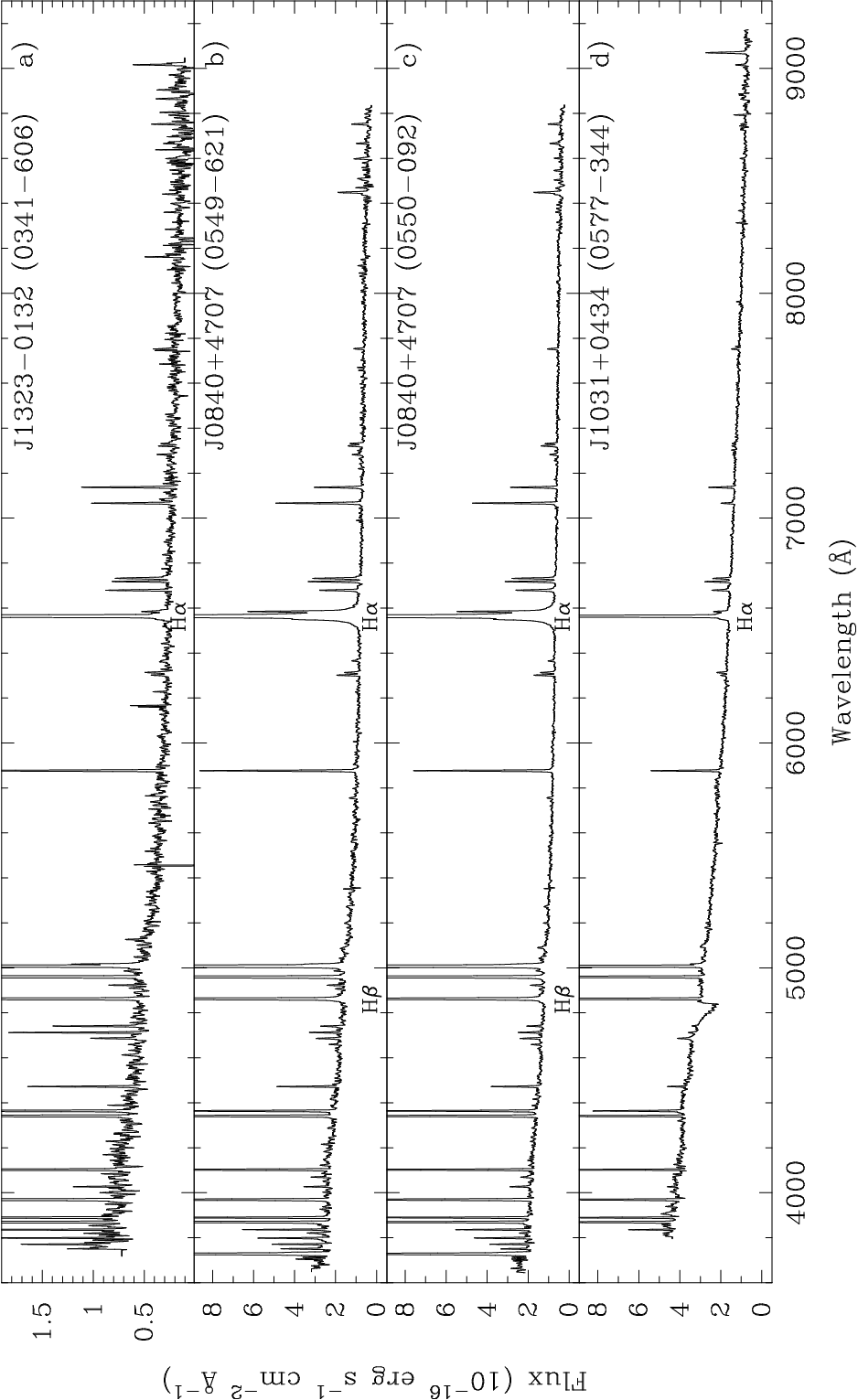}{1pt}{-90.}{370.}{520.}{-30.}{0.}
\figcaption{Redshift-corrected spectra of galaxies from SDSS subsample 1. 
Each galaxy is labeled by its SDSS name and, in parentheses, by
the SDSS spectrum number. Emission
lines with broad components are labeled. \label{figbr}}
\end{figure*}

\begin{figure*}
\figurenum{4}
\epsscale{0.9}
\plotone{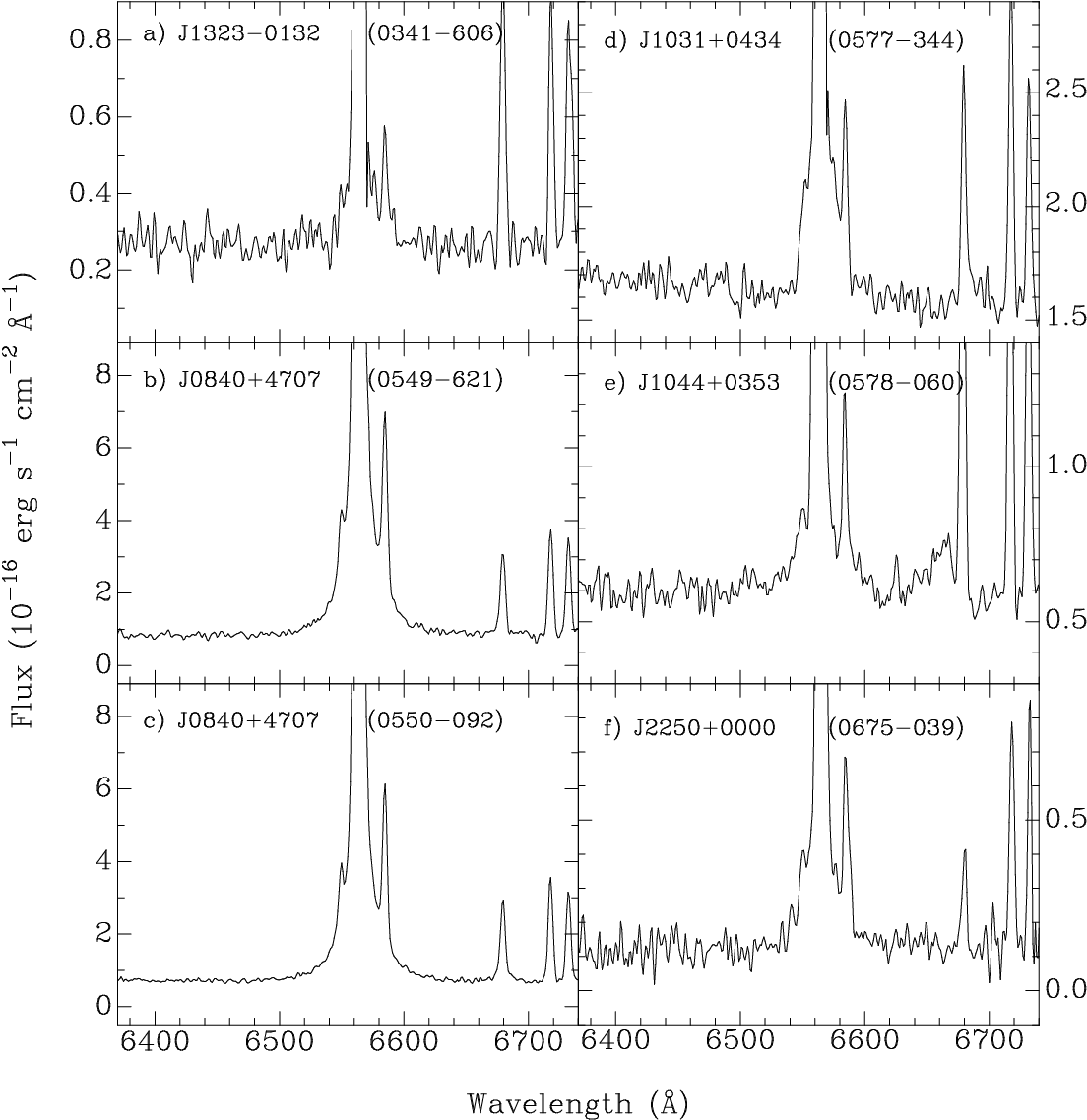}
\figcaption{The H$\alpha$ emission line in the spectra of SDSS subsample 1 
galaxies.\label{figbr1}}
\end{figure*}

\begin{figure*}
\figurenum{5}
\epsscale{1.1}
\plotfiddle{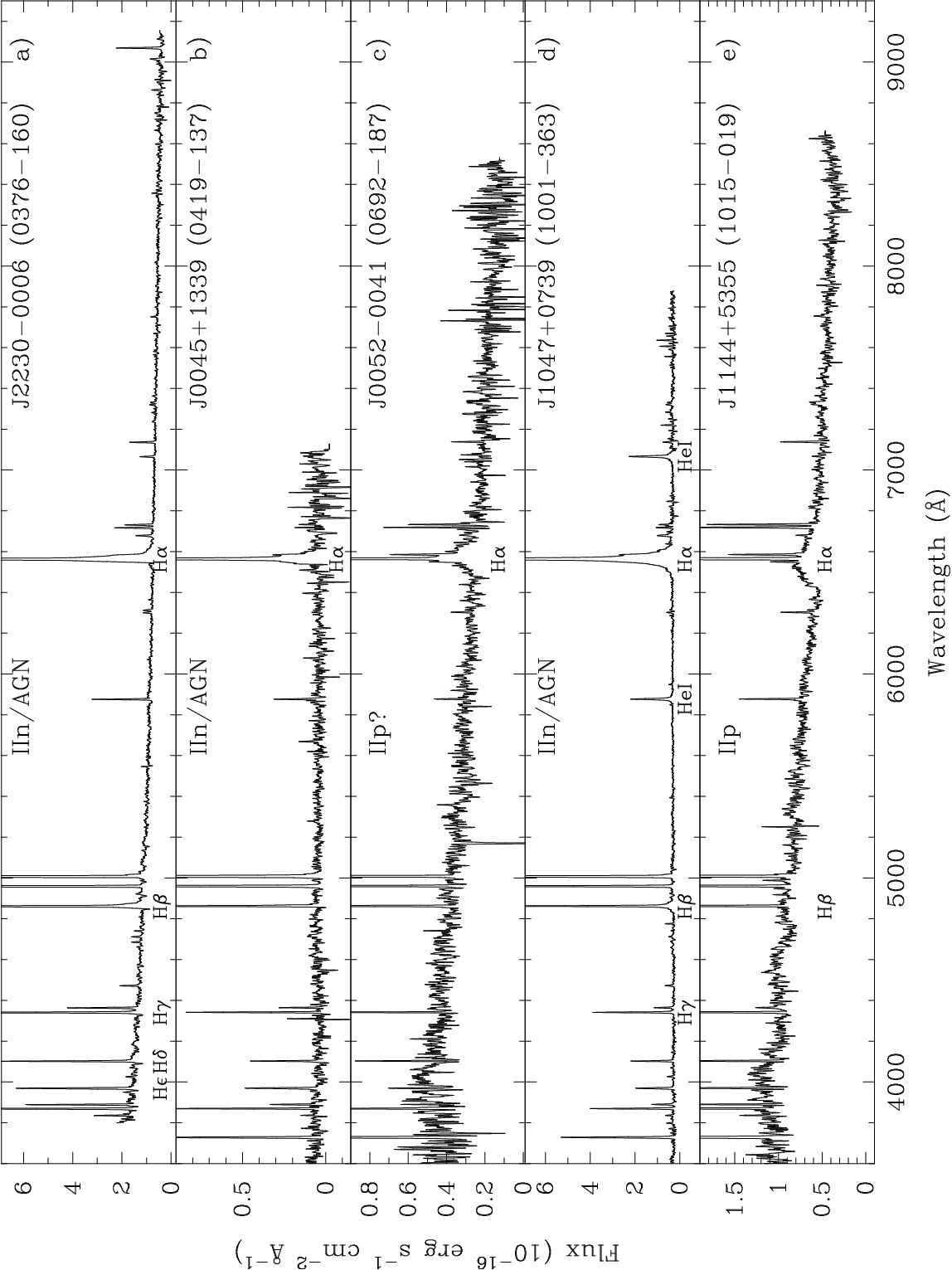}{1pt}{-90.}{370.}{520.}{-30.}{0.}
\figcaption{Redshift-corrected spectra of SDSS subsample 2 galaxies.
Each galaxy is labeled by its SDSS name and, in parentheses, by
the SDSS spectrum number. Emission
lines with broad components are labeled.\label{figSN}}
\end{figure*}

\begin{figure*}
\figurenum{6}
\epsscale{0.9}
\plotone{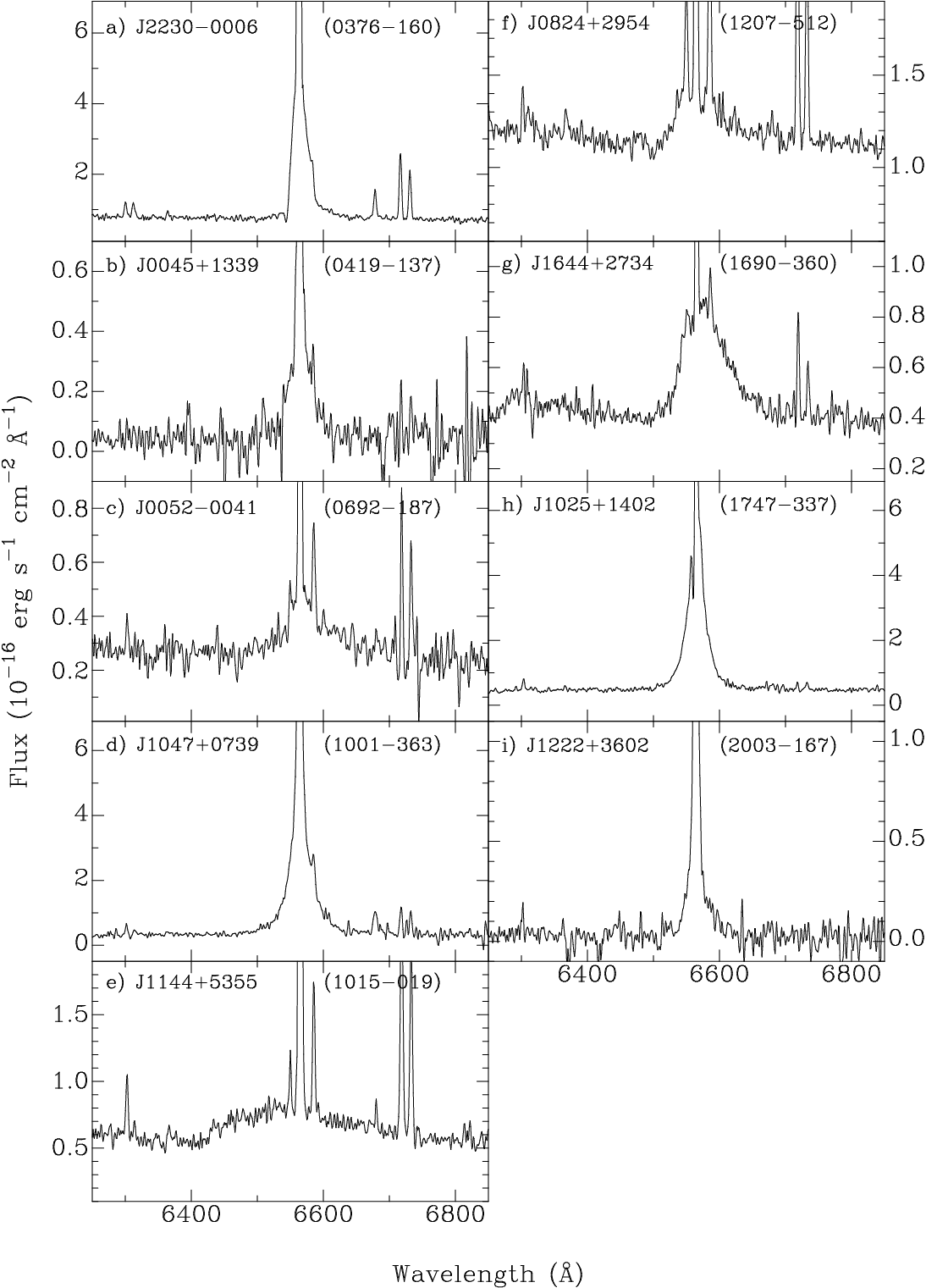}
\figcaption{The H$\alpha$ emission line in the spectra of SDSS subsample 2 
galaxies.\label{figSN1}}
\end{figure*}

In Fig. \ref{diagnDR5}, we show the location of the SDSS galaxies in the
[O {\sc iii}] $\lambda$5007/H$\beta$ -- [N {\sc ii}]
$\lambda$6583/H$\alpha$ diagnostic diagram. As said before, the dots denote 
the $\sim$ 10 000 emission-line galaxies with strong nebular lines,
presumably star-forming galaxies, that were selected from the SDSS DR5. 
Open circles show galaxies in SDSS subsample 1.  
Stars show those galaxies in SDSS subsample 2 that have spectra  
containing the broad features characteristic of type IIn 
SN/AGN spectra, and filled circles show those galaxies in SDSS subsample 2 
that have spectra containing the broad features characteristic 
of type IIp SN spectra.
It is seen that all of our 
sample galaxies lie in the star-forming galaxy region, far from
  the region occupied by solar and super-solar metallicity AGN, 
the location of which is indicated by ``AGN'' in Fig. \ref{diagnDR5}.
However, as we have already pointed
  out, models of low-metallicity NLR of AGN lie also in the same star-forming 
region of the diagram \citep{Gr06,S06}, so that we cannot distinguish 
between the two possibilities.

\citet{Gr06} have searched the SDSS for AGN with subsolar metallicity, using 
a variety of emission-line diagnostics and found 
only $\sim$ 40 clear candidates
out of $\sim$ 23000 Seyfert 2 galaxies. The SDSS spectra
of these selected subsolar metallicity AGNs \citep{Gr06} 
are however very different from those of our type 
IIn SN/AGN candidates. They show relatively strong He {\sc ii} $\lambda$4686,
[O {\sc i}] $\lambda$6300, and [S {\sc ii}] $\lambda$6717, $\lambda$6731
emission lines. They also show clear evidence for an 
  old underlying stellar population with strong stellar absorption
  features. On the other hand, our IIn SN/AGN candidates exhibit no clear
  evidence for an underlying old stellar population in their spectra, 
and are characterized by high equivalent widths of H$\beta$. It is therefore
  likely that our galaxies are more metal-poor than those 
of \citet{Gr06}. Additionally, the
  contribution of the star-forming regions to the total light is higher in our
  galaxies.

Thus, if we adopt the hypothesis that the broad emission 
in SDSS subsample 2 objects 
that do not possess type IIp SN characteristics is caused, not by 
type IIn SNe, but by an AGN, we would not have here 
the usual AGN phenomenon. These objects would form a new class of 
low-luminosity AGN that are extremely
rare. These AGN with intermediate-mass black holes
(masses $\sim$ 10$^3$-10$^5$ solar masses) 
would reside in low-metallicity dwarf galaxies. The  
oxygen abundance 12 + log O/H  of these dwarf galaxies 
would be in the range 7.36 -- 7.99, more than 5--26 times lower
than typical metallicities of ``normal'' AGN.

To decide whether type IIn SNe or AGN are responsible for the broad emission 
in these galaxies, monitoring of their spectral features on the
relatively long time scale of several years is necessary. If broad features
are produced by IIn type SNe, then we would expect the SN to
fade and the broad lines to disappear after a few years. On the other hand, 
broad lines produced by an AGN will last much longer. Additionally, higher
  signal-to-noise ratio spectra are necessary to put better
  constraints on the presence of the high-ionization [Ne {\sc v}]
  $\lambda$3426 and He {\sc ii} $\lambda$4686 emission lines, 
probable indicators of a source of hard non-thermal radiation. We may also 
resort to non-optical observations. Deep radio and X-ray observations of 
these objects can provide additional tests for whether or not they 
harbor an AGN. For example, X-ray variability would be a clear indicator of the 
presence of a compact object.

\section{CONCLUSIONS}

We study here broad line emission in star-forming dwarf galaxies.
To this end, we have assembled  
high signal-to-noise ratio MMT spectra of 22 H {\sc ii} 
regions in low-metallicity blue compact dwarf (BCD) galaxies and of one
H {\sc ii} region in the spiral galaxy M 101. Fourteen of these MMT spectra 
show broad emission. Additionally, we 
have also 
extracted 21 spectra (the majority of which are spectra of BCDs) 
from the SDSS DR5 
which show broad emission in the H$\alpha$ line and other emission lines.

We have arrived at the following conclusions:

1. The broad components of strong emission lines in BCDs 
are not due to instrumental
effects, but are caused by real dynamical processes related to the evolution
of massive stars. Broad luminosities in the range 10$^{36}$--10$^{39}$ 
erg s$^{-1}$ mostly originate 
in dense circumstellar envelopes of massive stars with stellar 
winds and/or in supernovae remnants.

2. The widths of the broad components imply expansion velocities of
$\sim$ 1000 -- 7000 km s$^{-1}$, in the range of stellar wind velocities 
or of expansion velocities of SN remnants.  

3. There is not a tight correlation between the presence 
of broad components in strong emission lines, such as H$\beta$, H$\alpha$,
[O {\sc iii}] $\lambda$4959, 5007\AA, and Wolf-Rayet broad emission. While
we cannot rule out that some part 
of the broad emission originates in the circumstellar
envelopes of WR stars, this cannot be the case for all galaxies because
WR stars are not always seen in galaxies
with broad H$\beta$ and H$\alpha$ emission. 

4. Steep Balmer decrements of broad hydrogen lines, 
high [O {\sc iii}] $\lambda$4363/($\lambda$4959 + $\lambda$5007) broad line
flux ratios and high fluxes of He {\sc i} $\lambda$5876 and especially
of He {\sc i} $\lambda$7065 emission lines 
in some spectra suggest that broad emission is
formed in very dense regions such as in 
hydrogen-rich circumstellar envelopes of 
massive Ofp/LBV stars, in young SN remnants and in the NLR of an AGN.

5. In several cases, the flux of the broad component is constant
over a period of several years. The luminosities of the broad lines
are lower than those in SNe, but are consistent with the luminosities of
emission lines expected from stellar winds of one to several dozens of  
Ofp/LBV stars. These properties are characteristic of galaxies in the MMT 
sample, 
where the flux ratio of broad to narrow components is of $\sim$ 1 -- 2\%, or
in the SDSS subsample 1, where the flux ratio of broad to narrow components 
does not exceed 0.25. On the other hand, SDSS subsample 2 contains 
9 galaxies with a broad to narrow component flux ratio larger than 0.25, 
and with the highest broad line luminosities.
Of these 9 objects, 4 have spectra of type IIp SN, while 
the spectra of the remaining 5 objects have been tentatively classified  
as of type IIn SN/AGN.
These IIn SN/AGN spectra have extremely high 
broad H$\alpha$ luminosities, in the range (2.7 -- 16.0)$\times$10$^{41}$
erg s$^{-1}$, placing them among the brightest IIn type SNe known.
These high luminosities 
are comparable to those of low-luminosity AGN, so that gas accretion onto 
an intermediate-mass black hole cannot be ruled out as the source of the 
broad emission in the objects with the highest broad line luminosities.  

6. There is no obvious spectroscopic evidence for the  
presence of a source of non-thermal hard ionizing radiation in all of our  
sample galaxies.
We cannot exclude however the presence of an AGN with the
intermediate-mass black hole ($\sim$ 10$^3$-10$^5$ solar masses) in 
those galaxies with type IIn SN/AGN spectral characteristics. 
If these objects do indeed harbor a low-luminosity AGN, 
they would form a new class of AGN that are extremely
rare. These AGN with the intermediate-mass black hole
would reside in low-metallicity dwarf galaxies, with an oxygen abundance 
much lower than that of a typical AGN.

\acknowledgements
We thank N.N.Chugai and R.A.Chevalier for illuminating discussions 
on the spectroscopic properties of supernovae and G.Stasi\'nska for
useful discussions on the spectroscopic properties of low-metallicity AGN.
Y.I.I. and N.G.G. are grateful to the staff of the Astronomy Department at the 
University of Virginia for their warm hospitality. 
We thank the financial support of National Science Foundation
grant AST02-05785. The MMT time was available thanks to a grant from the
Frank Levinson Fund of the Peninsula Community Foundation
to the Astronomy Department of the University of Virginia.
    Funding for the Sloan Digital Sky Survey (SDSS) and SDSS-II has been 
provided by the Alfred P. Sloan Foundation, the Participating Institutions, 
the National Science Foundation, the U.S. Department of Energy, the National 
Aeronautics and Space Administration, the Japanese Monbukagakusho, the 
Max Planck Society, and the Higher Education Funding Council for England. 


\clearpage

\begin{deluxetable}{llrrrrc}
  \tabletypesize{\scriptsize}
  \tablecolumns{12}
  \tablewidth{0pc}
  \tablecaption{Parameters of the broad emission lines in the MMT galaxies \label{tab1}}
  \tablehead{
  \colhead{Object}    & \colhead{Line}     &\colhead{FWHM} &\colhead{FWZI} &\colhead{$I_{br}$\tablenotemark{a}}&
  \colhead{$I_{br}/I_{nar}$}&\colhead{$L_{br}$      } \\
  \colhead{    }      &\colhead{    }     &\colhead{(\AA)}&\colhead{(\AA)}&\colhead{        }&
  \colhead{       (\%)     }&\colhead{(erg s$^{-1}$)}
}
  \startdata
SBS 0335--052E&4861 H$\beta$        & 24.3     & 64.5     & 12.19   & 1.18                &4.30$\times$10$^{38}$ \\
~~~~--''--    &4959 [O {\sc iii}]   & 22.7     & 44.3     & 11.72   & 1.06                &4.18$\times$10$^{38}$ \\
~~~~--''--    &5007 [O {\sc iii}]   & 22.1     & 59.1     & 31.40   & 0.96                &1.12$\times$10$^{39}$ \\
SDSSJ1404+5423&4861 H$\beta$        & 17.1     & 32.9     & 29.94   & 1.05                &1.61$\times$10$^{37}$\tablenotemark{b} \\
~~~~--''--    &4959 [O {\sc iii}]   & 14.8     & 42.3     & 81.56   & 1.30                &4.38$\times$10$^{37}$\tablenotemark{b} \\
~~~~--''--    &5007 [O {\sc iii}]   & 16.4     & 54.4     &302.97   & 1.02                &1.63$\times$10$^{38}$\tablenotemark{b} \\
Haro 3        &4861 H$\beta$        & 15.5     & 30.2     & 13.27   & 1.71                &2.52$\times$10$^{37}$ \\
~~~~--''--    &4959 [O {\sc iii}]   & 14.0     & 32.9     & 23.48   & 2.23                &4.46$\times$10$^{37}$ \\
\enddata
\tablenotetext{a}{In units 10$^{-16}$ erg s$^{-1}$ cm$^{-2}$.}
\tablenotetext{b}{At the distance $D$ = 6.7 Mpc \citep{Pi04}.}
\tablenotetext{c}{At the distance $D$ = 5.7 Mpc \citep{K02}.}
\tablenotetext{d}{At the distance $D$ = 3.42 Mpc \citep{TI05b}.}
  \end{deluxetable}

\clearpage

\begin{deluxetable}{llccccccl}
  \tabletypesize{\scriptsize}
  \tablecolumns{8}
  \tablewidth{0pc}
  \tablecaption{General characteristics of SDSS galaxies with broad emission
lines\label{tab2}}
  \tablehead{
\colhead{SDSS Spectrum ID}    &   \colhead{Galaxy}    &\colhead{R.A.}     &\colhead{DEC.} 
&\colhead{Redshift} &\colhead{$g$}&\colhead{$M_g$}&\colhead{12+logO/H}
& \colhead{Date of Obs.} }
  \startdata
\multicolumn{9}{c}{a) galaxies from the subsample 1} \\
51690-0341-606&SDSSJ1323$-$0132&13 23 47.5&$-$01 32 52&0.02254&18.15&--16.62&7.78& 26 May 2000 \\
51981-0549-621&SDSSJ0840$+$4707&08 40 29.9&$+$47 07 10&0.04219&17.60&--18.53&7.63& 13 Mar 2001 \\
51959-0550-092&SDSSJ0840$+$4707&08 40 29.9&$+$47 07 10&0.04220&17.60&--18.53&7.63& 19 Feb 2001 \\
52367-0577-344&SDSSJ1031$+$0434&10 31 37.3&$+$04 34 22&0.00401&16.03&--15.00&7.70& 02 Apr 2002 \\
52339-0578-060&SDSSJ1044$+$0353&10 44 57.8&$+$03 53 13&0.01309&17.48&--16.12&7.46& 06 Mar 2002 \\
52590-0675-039&SDSSJ2250$+$0000&22 50 59.3&$+$00 00 33&0.08078&19.77&--17.78&7.61& 10 Nov 2002 \\
52178-0676-192&SDSSJ2250$+$0000&22 50 59.3&$+$00 00 33&0.08078&19.77&--17.78&7.61& 26 Sep 2001 \\
\enddata
\tablecomments{J1402$+$5420 and J1404$+$5423 are H {\sc ii} regions in the
spiral galaxy M 101; J1323$-$0132 $\equiv$ UM 570; 
J0840$+$4707 $\equiv$ HS 0837+4717; J1536$+$3122 $\equiv$ CG 1308.}
  \end{deluxetable}

\clearpage

  \begin{deluxetable}{lrrrrr}
  \tabletypesize{\scriptsize}
  \tablecolumns{6}
  \tablewidth{0pc}
  \tablecaption{Emission line intensities of galaxies in the SDSS subsample 1
\label{tab3}}
  \tablehead{
  \colhead{\sc Ion}
  &\colhead{$I$($\lambda$)/$I$(H$\beta$)}
  &\colhead{$I$($\lambda$)/$I$(H$\beta$)}
  &\colhead{$I$($\lambda$)/$I$(H$\beta$)}
  &\colhead{$I$($\lambda$)/$I$(H$\beta$)}
  &\colhead{$I$($\lambda$)/$I$(H$\beta$)}
}
  \startdata
  & \multicolumn{5}{c}{Galaxy} \\ 
  & \multicolumn{5}{c}{(Spectrum)} \\ \cline{2-6}
  &
 \multicolumn{1}{c}{J1323$-$0132 }&
 \multicolumn{1}{c}{J0840$+$4707 }&
 \multicolumn{1}{c}{J1031$+$0434 }&
 \multicolumn{1}{c}{J1044$+$0353 }&
 \multicolumn{1}{c}{J2250$+$0000 } \\ 
  &
 \multicolumn{1}{c}{(0341-606) }&
 \multicolumn{1}{c}{(0549-621) }&
 \multicolumn{1}{c}{(0577-344) }&
 \multicolumn{1}{c}{(0578-060) }&
 \multicolumn{1}{c}{(0675-039) } \\ \tableline
3727 [O {\sc ii}]                 &   \nodata~~~~       &  46.47 $\pm$   1.66 &   \nodata~~~~       &   \nodata~~~~       &  57.15 $\pm$   3.64 \\
3750 H12                          &   4.70 $\pm$   1.52 &   4.27 $\pm$   0.45 &   \nodata~~~~       &   \nodata~~~~       &   5.31 $\pm$   2.27 \\
3771 H11                          &   6.25 $\pm$   1.06 &   5.07 $\pm$   0.45 &   \nodata~~~~       &   \nodata~~~~       &   6.77 $\pm$   2.00 \\
3798 H10                          &   7.03 $\pm$   0.91 &   6.03 $\pm$   0.47 &   \nodata~~~~       &   6.24 $\pm$   0.53 &   7.89 $\pm$   1.89 \\
\enddata
\tablenotetext{a}{in units 10$^{-14}$ erg s$^{-1}$ cm$^{-2}$.}
  \end{deluxetable}

\clearpage

  \begin{deluxetable}{lrrrrr}
  \tabletypesize{\scriptsize}
  \tablecolumns{6}
  \tablewidth{0pc}
  \tablecaption{Emission line intensities of galaxies in the SDSS subsample 2
\label{tab4}}
  \tablehead{
  \colhead{\sc Ion}
  &\colhead{$I$($\lambda$)/$I$(H$\beta$)}
  &\colhead{$I$($\lambda$)/$I$(H$\beta$)}
  &\colhead{$I$($\lambda$)/$I$(H$\beta$)}
  &\colhead{$I$($\lambda$)/$I$(H$\beta$)}
  &\colhead{$I$($\lambda$)/$I$(H$\beta$)}}
  \startdata
  & \multicolumn{5}{c}{Galaxy} \\ 
  & \multicolumn{5}{c}{(Spectrum)} \\ \cline{2-6}
  &
 \multicolumn{1}{c}{J2230$-$0006 }&
 \multicolumn{1}{c}{J0045$+$1339 }&
 \multicolumn{1}{c}{J0052$-$0041 }&
 \multicolumn{1}{c}{J1047$+$0739 }&
 \multicolumn{1}{c}{J1144$+$5355 } \\ 
  &
 \multicolumn{1}{c}{(0376-160) }&
 \multicolumn{1}{c}{(0419-137) }&
 \multicolumn{1}{c}{(0692-187) }&
 \multicolumn{1}{c}{(1001-363) }&
 \multicolumn{1}{c}{(1015-019) } \\  \tableline
3727 [O {\sc ii}]                 &   \nodata~~~~       &  75.04 $\pm$   7.00 & 262.29 $\pm$  22.00 & 101.96 $\pm$   5.40 & 227.99 $\pm$  11.72 \\
3771 H11                          &   \nodata~~~~       &   \nodata~~~~       &   \nodata~~~~       &   4.60 $\pm$   1.60 &   \nodata~~~~       \\
3798 H10                          &   \nodata~~~~       &   \nodata~~~~       &   \nodata~~~~       &   5.27 $\pm$   1.85 &   \nodata~~~~       \\
3835 H9                           &   8.70 $\pm$   1.23 &   \nodata~~~~       &   \nodata~~~~       &   6.23 $\pm$   1.82 &  12.14 $\pm$   3.05 \\
\enddata
\tablenotetext{a}{in units 10$^{-14}$ erg s$^{-1}$ cm$^{-2}$.}
  \end{deluxetable}

\clearpage

  \begin{deluxetable}{lrrrrr}
  \tabletypesize{\scriptsize}
  \tablecolumns{6}
  \tablewidth{0pc}
  \tablecaption{ Ionic and total heavy element abundances of galaxies 
in the SDSS subsample 1 \label{tab5}}
  \tablehead{
  \colhead{Property}&\colhead{Value}&\colhead{Value}&\colhead{Value}&\colhead{Value}&\colhead{Value}}
  \startdata
  & \multicolumn{5}{c}{Galaxy} \\ 
  &\multicolumn{5}{c}{(Spectrum)} \\ \cline{2-6}
  &
 \multicolumn{1}{c}{J1323$-$0132 }&
 \multicolumn{1}{c}{J0840$+$4707 }&
 \multicolumn{1}{c}{J1031$+$0434 }&
 \multicolumn{1}{c}{J1044$+$0353 }&
 \multicolumn{1}{c}{J2250$+$0000 } \\ 
  &
\multicolumn{1}{c}{(0341-606)} & \multicolumn{1}{c}{(0549-621)} &
\multicolumn{1}{c}{(0577-344)} & \multicolumn{1}{c}{(0578-060)} &
\multicolumn{1}{c}{(0675-039)} \\ \tableline
  $T_{\rm e}$(O {\sc iii}) (K) &
$17397  \pm 511 $ & $18767  \pm 497 $ &
$16679  \pm 618 $ & $19441  \pm 557 $ &
$18758  \pm1057 $ \\

  $T_{\rm e}$(O {\sc ii}) (K) &
$15082  \pm 413 $ & $15444  \pm 380 $ &
$14818  \pm 512 $ & $15554  \pm 413 $ &
$15442  \pm 809 $ \\

  $T_{\rm e}$(S {\sc iii}) (K) &
$16045  \pm 424 $ & $17571  \pm 413 $ &
$15877  \pm 513 $ & $18194  \pm 462 $ &
$17644  \pm 877 $ \\

  $N_{\rm e}$(S {\sc ii}) (cm$^{-3}$) &
$ 1420  \pm1010 $ & $  513  \pm 224 $ &
$   10  \pm  10 $ & $  302  \pm 232 $ &
$  114  \pm 240 $ \\
\enddata
  \end{deluxetable}

\clearpage

  \begin{deluxetable}{lrrrrr}
  \tabletypesize{\scriptsize}
  \tablecolumns{6}
  \tablewidth{0pt}
  \tablecaption{Ionic and total heavy element abundances of galaxies 
in the SDSS subsample 2 \label{tab6}}
  \tablehead{
  \colhead{\sc Property}&\colhead{Value}&\colhead{Value}&\colhead{Value}&\colhead{Value}&\colhead{Value}}
  \startdata
  & \multicolumn{5}{c}{Galaxy} \\ 
 &\multicolumn{5}{c}{(Spectrum)} \\ \cline{2-6}
  &
 \multicolumn{1}{c}{J2230$-$0006 }&
 \multicolumn{1}{c}{J0045$+$1339 }&
 \multicolumn{1}{c}{J0052$-$0041 }&
 \multicolumn{1}{c}{J1047$+$0739 }&
 \multicolumn{1}{c}{J1144$+$5355 } \\ 
 &
\multicolumn{1}{c}{(0376-160)} & \multicolumn{1}{c}{(0419-137)} &
\multicolumn{1}{c}{(0692-187)} & \multicolumn{1}{c}{(1001-363)} &
\multicolumn{1}{c}{(1015-019)} \\ \tableline
  $T_{\rm e}$(O {\sc iii}) (K) &
$17115  \pm 665 $ & $14641  \pm1245 $ &
$10836  \pm3598 $ & $14392  \pm 713 $ &
$12629  \pm1484 $ \\

  $T_{\rm e}$(O {\sc ii}) (K) &
$14985  \pm 543 $ & $13789  \pm1094 $ &
$10884  \pm3421 $ & $13635  \pm 631 $ &
$12366  \pm1366 $ \\

  $T_{\rm e}$(S {\sc iii}) (K) &
$16324  \pm 552 $ & $13488  \pm1033 $ &
$10876  \pm2986 $ & $13159  \pm 592 $ &
$12150  \pm1232 $ \\

  $N_{\rm e}$(S {\sc ii}) (cm$^{-3}$) &
$   48  \pm 137 $ & $   10  \pm  10 $ &
$  300  \pm 482 $ & $  181  \pm 265 $ &
$   16  \pm 100 $ \\
\enddata
  \end{deluxetable}

\clearpage

\begin{deluxetable}{lllrrrrc}
  \tabletypesize{\scriptsize}
  \tablecolumns{12}
  \tablewidth{0pc}
  \tablecaption{Parameters of the broad emission lines of the SDSS 
subsample 1 galaxies \label{tab7}}
  \tablehead{
  \colhead{Galaxy}    &  \colhead{Spectrum}    & \colhead{Line}     &\colhead{FWHM} &\colhead{FWZI} &\colhead{$I_{br}$}&
  \colhead{$I_{br}/I_{nar}$}&\colhead{$L_{br}$      } \\
  \colhead{    }        &\colhead{    }      &\colhead{    }     &\colhead{(\AA)}&\colhead{(\AA)}&\colhead{        }&
  \colhead{       (\%)     }&\colhead{(erg s$^{-1}$)} 
}
  \startdata
J1323$-$0132&0341-606  &6563 H$\alpha$        & 28.5     & 40.5     &  7.34$\pm$2.25   & 1.80                &7.14$\times$10$^{38}$ \\
J0840$+$4707&0549-621  &4861 H$\beta$         & 16.5     & 35.5     & 31.01$\pm$4.75   & 6.03                &1.05$\times$10$^{40}$ \\
~~~~--''--&~~~~--''--&6563 H$\alpha$        & 29.1     &147.4     &244.81$\pm$7.47   &17.21                &8.35$\times$10$^{40}$ \\
~~~~--''--&0550-092  &6563 H$\alpha$        & 26.7     &153.7     &232.56$\pm$7.31   &19.37                &7.92$\times$10$^{40}$ \\
J1031$+$0434&0577-344  &6563 H$\alpha$        & 25.1     & 41.1     & 24.72$\pm$4.30   & 5.34                &7.60$\times$10$^{37}$ \\
J1044$+$0353&0578-060  &6563 H$\alpha$        & 41.6     & 89.5     & 16.80$\pm$4.71   & 1.34                &5.51$\times$10$^{38}$ \\
J2250$+$0000&0675-039  &6563 H$\alpha$        & 28.7     & 50.0     & 23.27$\pm$3.14   & 7.61                &2.91$\times$10$^{40}$ \\
~~~~--''--&0676-192  &6563 H$\alpha$        & 28.5     & 52.5     & 25.15$\pm$3.43   & 7.55                &3.14$\times$10$^{40}$ \\
\enddata
\tablenotetext{a}{At the distance $D$ = 6.7 Mpc \citep{Pi04}.}
  \end{deluxetable}

\clearpage

\begin{deluxetable}{lllrrrrc}
  \tabletypesize{\scriptsize}
  \tablecolumns{8}
  \tablewidth{0pc}
  \tablecaption{Parameters of the broad emission lines of the SDSS 
subsample 2 galaxies \label{tab8}}
  \tablehead{
  \colhead{Galaxy}    &  \colhead{Spectrum}    & \colhead{Line}     &\colhead{FWHM} &\colhead{FWZI} &\colhead{$I_{br}$}&
  \colhead{$I_{br}/I_{nar}$}&\colhead{$L_{br}$      } \\
  \colhead{    }      &  \colhead{    }      &\colhead{    }     &\colhead{(\AA)}&\colhead{(\AA)}&\colhead{        }&
  \colhead{       (\%)     }&\colhead{(erg s$^{-1}$)} 
}
  \startdata
J2230$-$0006&0376-160  &4340 H$\gamma$        & 12.2     &  22.4    & 12.92$\pm$4.54   & 13.76               &7.74$\times$10$^{37}$ \\
~~~~--''--&~~~~--''--&4861 H$\beta$         & 15.5     &  42.1    & 45.74$\pm$4.47   & 22.66               &2.74$\times$10$^{38}$ \\
~~~~--''--&~~~~--''--&6563 H$\alpha$        & 22.6     & 128.0    &144.14$\pm$6.31   & 25.53               &8.62$\times$10$^{38}$ \\
J0045$+$1339&0419-137  &6563 H$\alpha$        & 37.4     & 127.4    & 16.38$\pm$1.74   & 41.65               &2.74$\times$10$^{41}$ \\
J0052$-$0041&0692-187  &6563 H$\alpha$        & 77.8     & 141.6    & 20.42$\pm$3.61   & 49.15               &2.45$\times$10$^{40}$ \\
J1047$+$0739&1001-363  &4340 H$\gamma$        & 12.2     &  24.3    &  7.66$\pm$3.07   & 19.86               &4.16$\times$10$^{40}$ \\
~~~~--''--&~~~~--''--&4861 H$\beta$         & 25.2     &  65.5    & 38.31$\pm$4.51   & 45.83               &2.08$\times$10$^{41}$ \\
~~~~--''--&~~~~--''--&5876 He {\sc i}       & 18.1     &  45.5    & 10.33$\pm$2.07   & 40.32               &5.60$\times$10$^{40}$ \\
\enddata
  \end{deluxetable}

\clearpage

\clearpage

\epsscale{0.9}
{\plotone{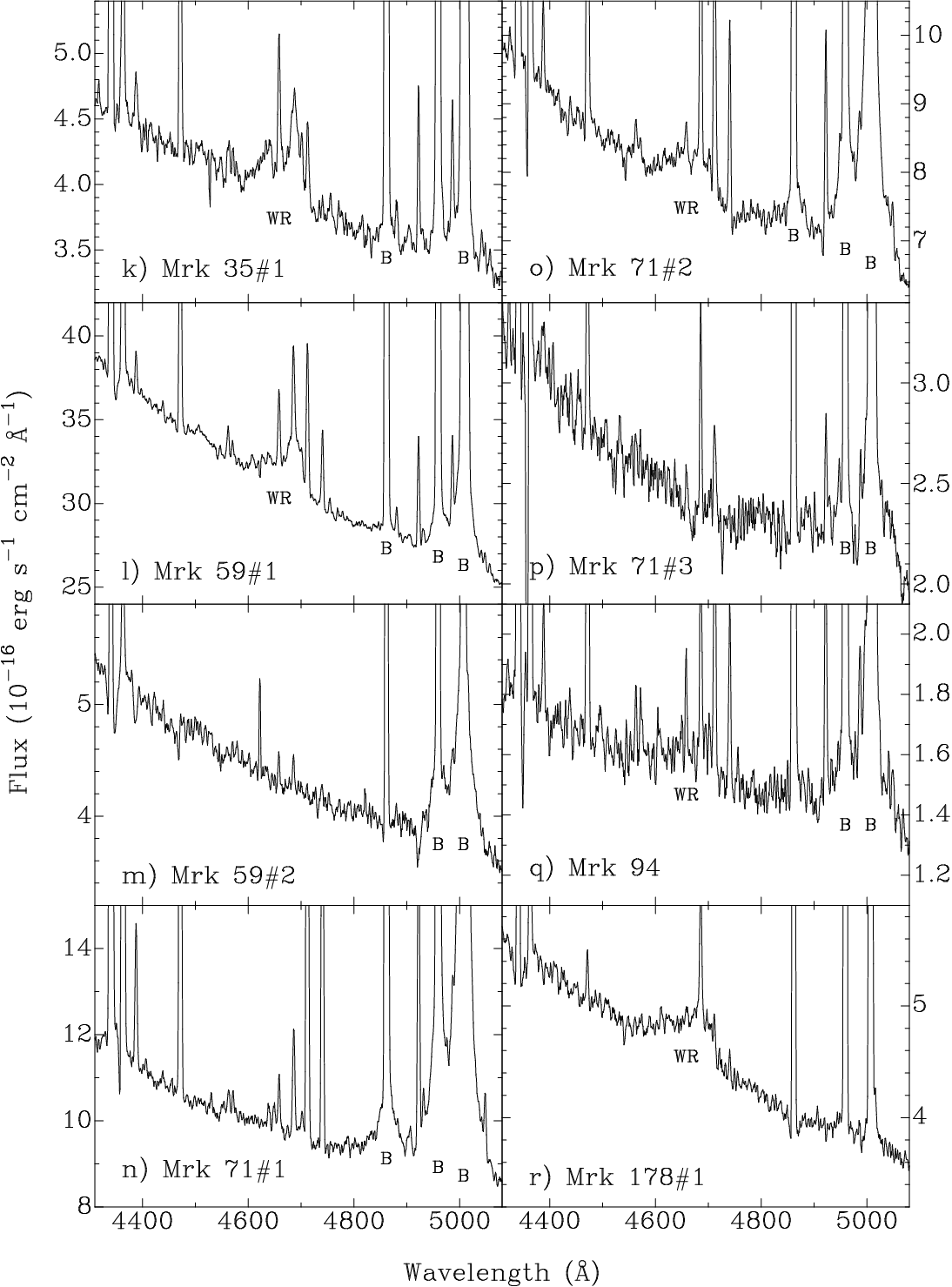}}\\
\centerline{Fig. 1. --- Continued.}

\clearpage

\epsscale{0.9}
{\plotone{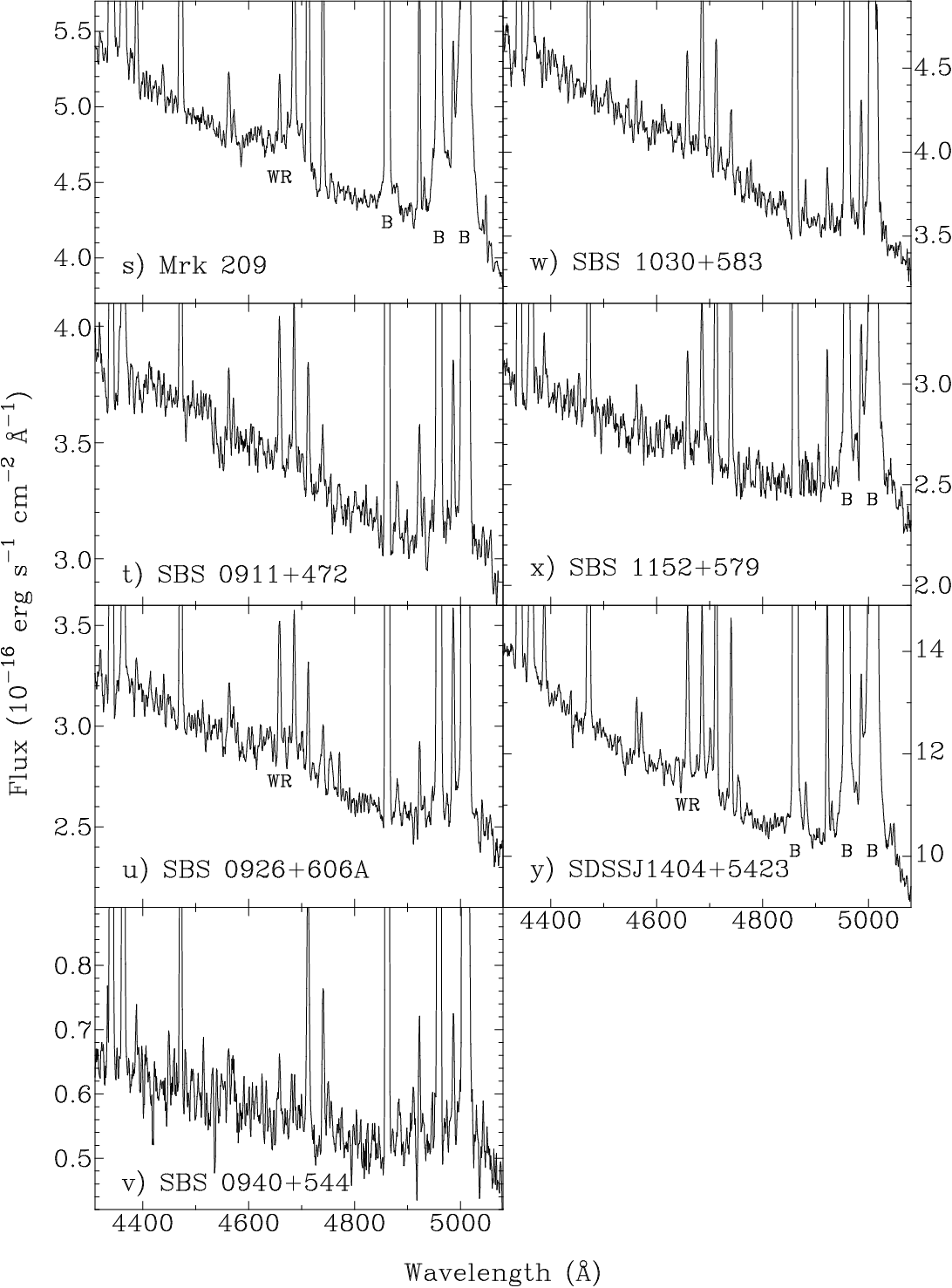}}\\
\centerline{Fig. 1. --- Continued.}

\clearpage

\clearpage

\clearpage

\epsscale{1.1}
{\plotfiddle{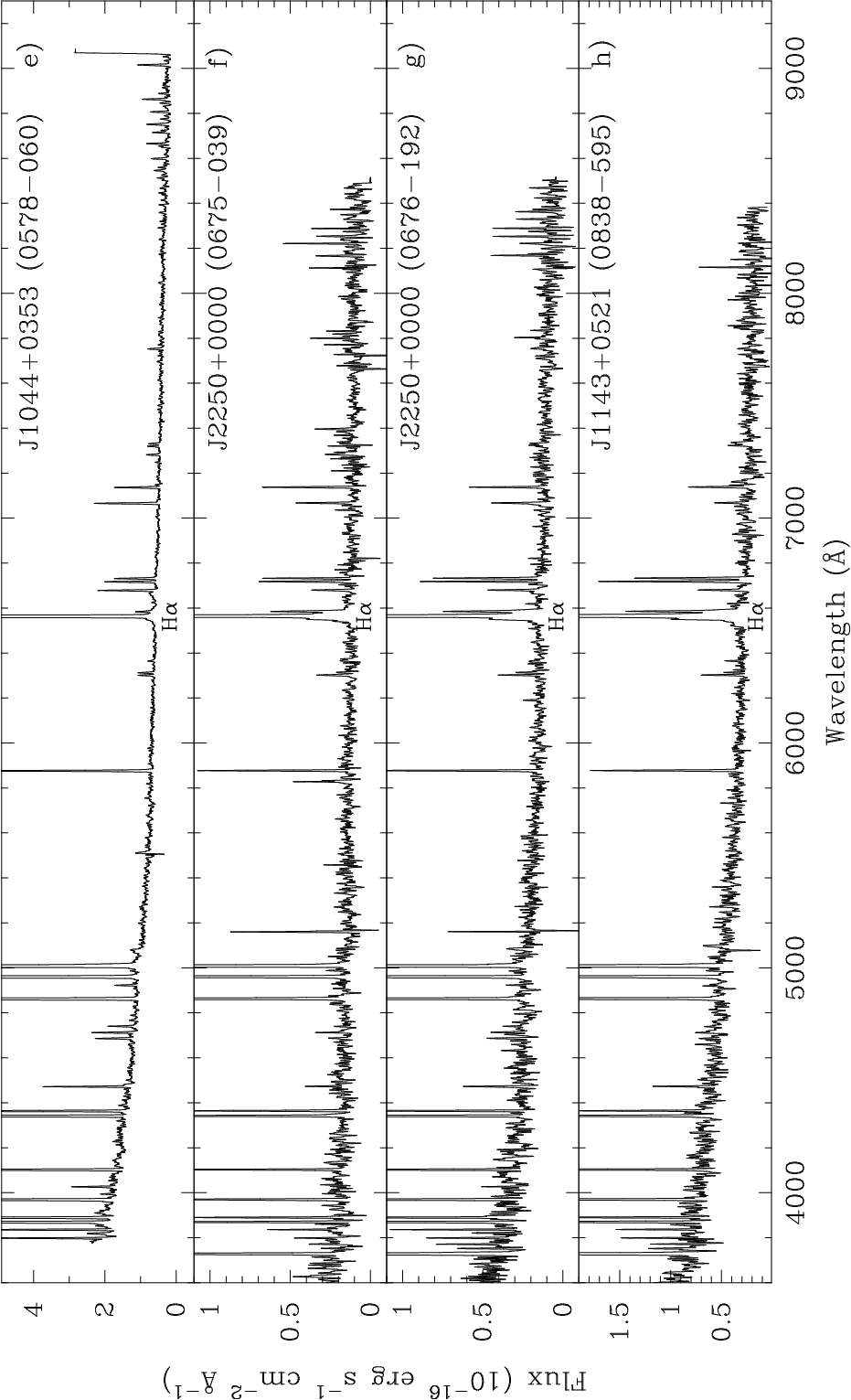}{1pt}{-90.}{370.}{520.}{-30.}{0.}}
\centerline{Fig. 3. --- Continued.}

\clearpage

\epsscale{1.1}
{\plotfiddle{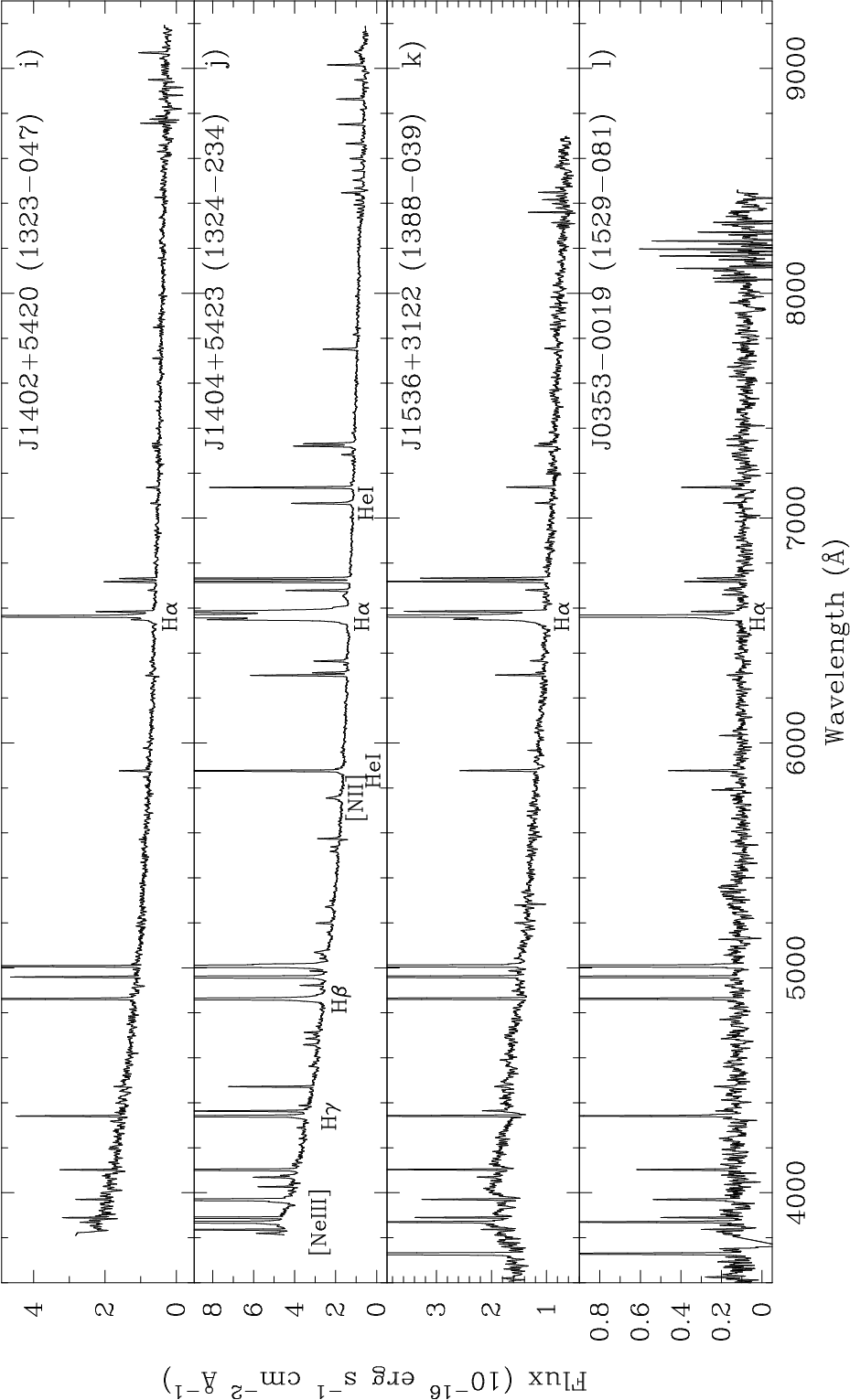}{1pt}{-90.}{370.}{520.}{-30.}{0.}}
\centerline{Fig. 3. --- Continued.}

\clearpage

\clearpage

\epsscale{0.9}
{\plotone{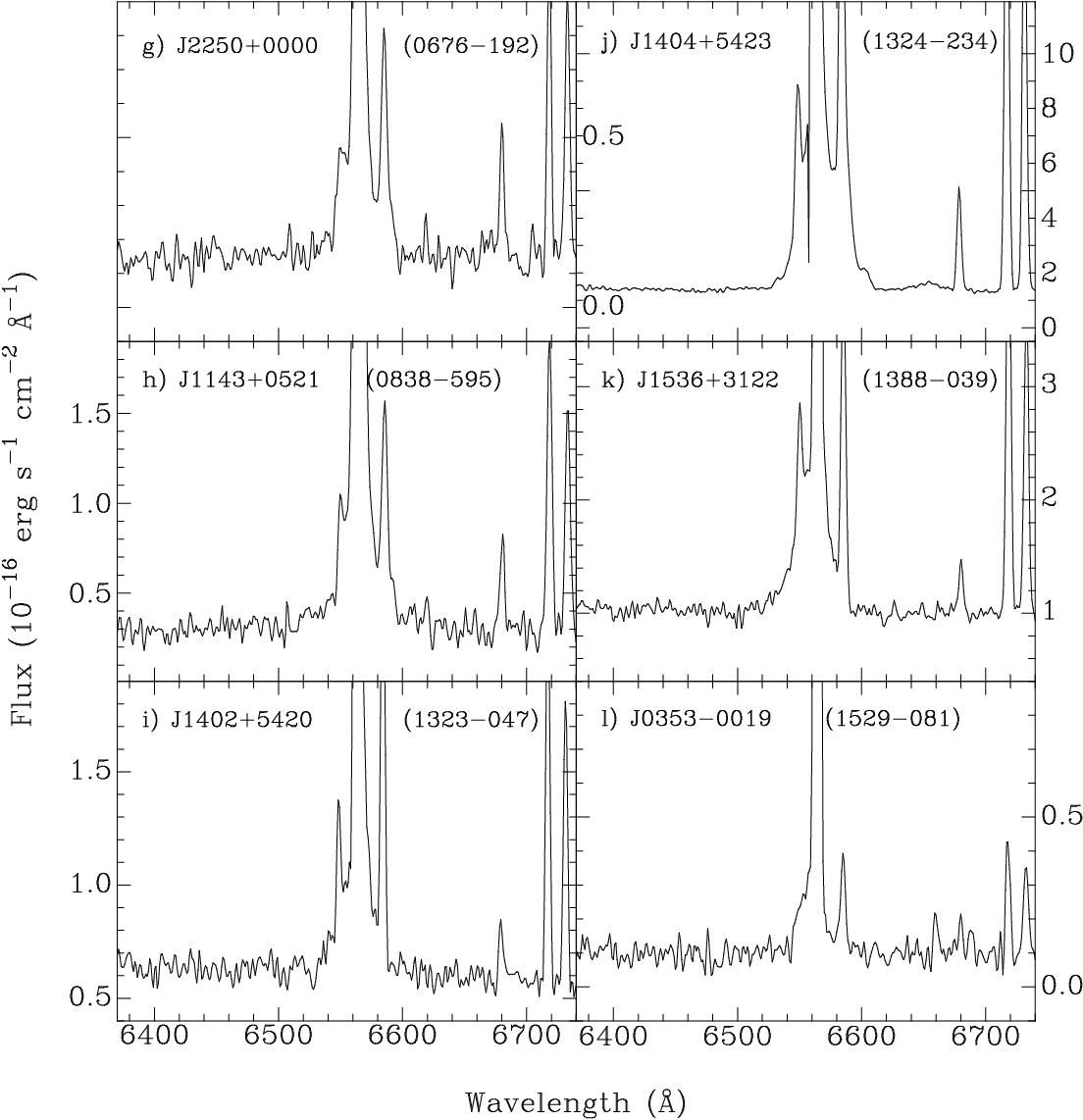}}\\
\centerline{Fig. 4. --- Continued.}

\clearpage

\clearpage

\epsscale{1.1}
{\plotfiddle{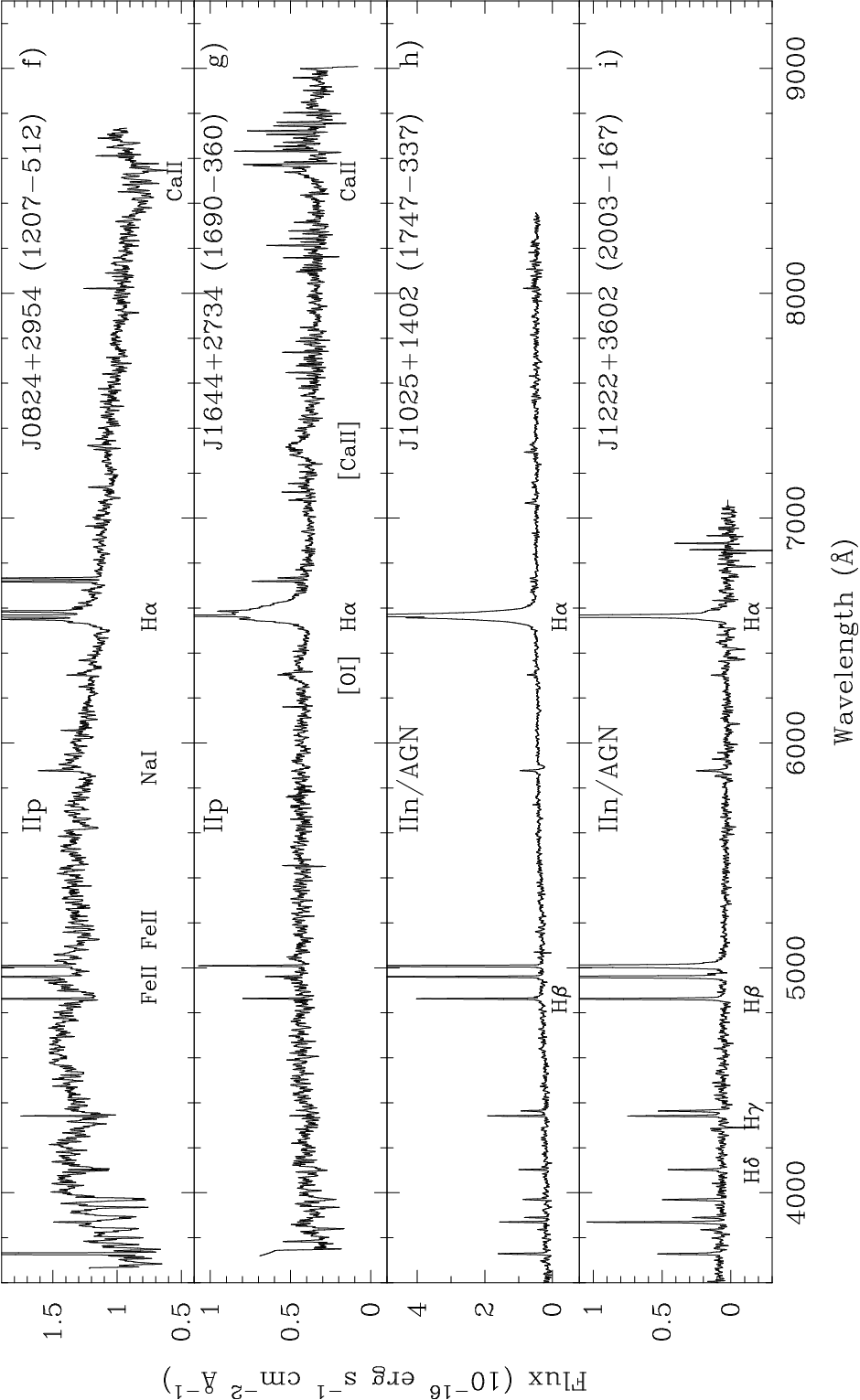}{1pt}{-90.}{370.}{520.}{-30.}{0.}}\\
\centerline{Fig. 5. --- Continued.}

\clearpage

\end{document}